\documentclass[aps,prb,twocolumn,showpacs]{revtex4}
\usepackage{graphicx}
\usepackage{amsmath}

\begin{document}

\title{Orbital Magnetic Susceptibility of Disordered Mesoscopic Systems}
\author{Moshe Goldstein}
\author{Richard Berkovits}
\affiliation{The Minerva Center, Department of Physics, Bar-Ilan University,
  Ramat-Gan 52900, Israel}

\begin{abstract}
In this paper we study the orbital weak-field susceptibility of
two-dimensional diffusive mesoscopic systems. For the previously
unstudied regime of temperatures lower than the mean level
spacing we find unexpected strong temperature as well as
statistical ensemble dependence of the average and typical
susceptibilities. An explanation for these features is given in
terms of the long tail of the zero-temperature susceptibility
distribution, including the parametric form of the temperature
dependence. For temperatures higher than the mean level spacing
we calculate the difference between the true canonical ensemble
and the effective grand-canonical ensemble. We also perform
numerical simulations, which seem to generally confirm previous
theoretical predictions for this regime of temperatures, although
some difficulties arise. The important role of gauge-invariance,
especially how it renders Random Matrix Theory inapplicable to the
study of orbital susceptibility, is discussed. We conclude by
considering interaction effects, giving a new interpretation to
previous results as well as demonstrating the influence of
in-plane magnetic field on the interaction-induced orbital response.
\end{abstract}


\maketitle

\section{Introduction}

Orbital magnetism is a purely quantum phenomenon, completely absent in
classical physics, due to a theorem of Born and van-Leeuwen
\cite{vanleeuwen}. It is therefore not surprising that the magnetic
response of mesoscopic systems, in which the particles move
phase-coherently across the sample, shows many peculiar features, and
is generally much larger than the magnetic response of macroscopic
systems.

Although this issue has been the subject of an intense scrutiny for
almost two decades, several issues remain obscure. Since the problem was
examined almost exclusively using diagrammatic perturbation expansion
for diffusive systems and semiclassical analysis for ballistic ones,
the behavior of the susceptibility at temperatures lower than the mean
level spacing was not adequately understood. The consideration of the
role of the different statistical ensembles is also incomplete, as we
discuss below, even for temperatures higher than the mean level
spacing. Moreover, although much numerical effort was concentrated on
investigating the magnetic response of multiply-connected geometries
(persistent current), no similar effort was directed into understanding
diffusive singly-connected geometries. Our aim in this paper is to
give some theoretical considerations for the above mentioned issues,
and also to compare these and previous results with numerical
simulations.

We will start with a survey of previous results. Here, as well as in
our analytical and numerical calculations, we limit ourselves to the
case of singly-connected two dimensional systems of spinless
noninteracting electrons (except in section~\ref{sec:v}), and to the
linear response regime.  We will always consider the susceptibility per
unit area.

The orbital magnetic response of macroscopic systems is the celebrated
Landau diamagnetic susceptibility \cite{landau},
$\chi_{{}_\mathrm{L}}=-e^2 / 24\pi mc^2$ . This form is
restricted to the linear response regime, $\hbar \omega_c < T$, where
$T$ is the temperature (we use throughout units where the Boltzmann
constant is unity) and $\omega_c=eH/mc$ is the cyclotron frequency (H
denotes the applied magnetic field). The Landau susceptibility is
temperature independent, up to temperatures of the order of the Fermi
energy $\epsilon_F$, above which the above result is multiplied by
$\epsilon_F/T$.

This macroscopic Landau susceptibility is quite small. In a
semiclassical picture this is a consequence of the absence of
flux-enclosing trajectories in infinite clean systems. However, in
mesoscopic systems such trajectories do exist, due to scattering by
either impurities or the sample boundaries. These two mechanisms tend
to strongly enhance the susceptibility as described below.
In the following we will assume that the dephasing caused by inelastic
scattering (by other electrons, phonons, etc.) is negligible, i.e.,
$L_{\phi} > L$, where $L_{\phi}$ is the phase coherence length,
and $L$ is the system's size.

We will first consider disorder-induced susceptibility, which is
dominant in the diffusive regime, $k_F^{-1} < \ell < L$, where $k_F$
is the Fermi wave-vector and $\ell$ is the elastic mean free path.
The presence of disorder causes large sample specific fluctuations.
Thus, the typical susceptibility, measured by the r.m.s. value
$\delta \chi$, is much larger than the average susceptibility, where
the average is over an ensemble of disorder realizations. For
temperatures below the Thouless energy
$E_C=\hbar D / L^2$ ($D=v_F \ell / 2$ is the diffusion coefficient
in two dimensions), the typical susceptibility is expected to vary as
$|\chi_{{}_\mathrm{L}}|(k_F \ell) \left[\ln(E_C / T)\right]^{1/2}$.
The form of the
logarithmic factor is correct in the very low field regime
$\Phi < {(T / E_C)}^{1/2}\Phi_0$, where $\Phi=HS$ is the total
magnetic flux through the system (S is the system's area), and
$\Phi_0=\hbar c / e$ is the flux quantum\cite{serota91}. These
fluctuations may persist even in temperatures above the Thouless
energy, where the temperature dependent factor changes to
$(E_C / T)^{1/2}$, with an increased range of validity for linear
response \cite{shapiro99}.

We now regard the disorder averaged susceptibility. Here we should
distinguish between the Canonical Ensemble (CE) and the Grand
Canonical Ensemble (GCE). The GCE corresponds to a situation where
the systems in the ensemble of disorder realizations can exchange
particles with a heat
bath with a fixed, system independent chemical potential. The CE
corresponds to a situation where the systems are isolated, with
the same fixed number of particles in each. Usually it was assumed
that the CE can be described by an Equivalent Grand Canonical Ensemble
(EGCE), in which each system can exchange particles with a different
heat bath, with its chemical potential adjusted to the realization
so that the thermally averaged number of particles in the system equals
the given fixed particle number. Thus, the difference appears only
after averaging over an ensemble of systems with different disorder
realizations, but not for a single system. However, it was
argued several times in the past that the CE and EGCE are really
equivalent only at zero temperature or at high
temperatures (higher than the mean level spacing $\Delta$ or even the
Thouless energy $E_C$). For intermediate temperatures the level
occupancy in the CE cannot be described by the Fermi-Dirac
distribution \cite{denton73, kamenev97}.

It is predicted that in the GCE the ensemble averaged susceptibility
will be equal to the Landau diamagnetic response. Disorder only
modifies this result by a small contribution, of order
${(k_F \ell)}^{-1}$. On the other hand, in the EGCE there is an
additional paramagnetic contribution \cite{serota91,altshuler93}, of
order $|\chi_{{}_\mathrm{L}}|(k_F \ell) \Delta / T$.
This contribution is much larger than the Landau susceptibility, and
will dominate both the sign and the magnitude of the response up to
temperatures of the order of the Thouless energy $E_C$. This
paramagnetic part of the response is expected to be strongly field
dependent even in the weak field regime $\Phi < \Phi_0$ , and linear
response is expected to apply only for very low fields, for which
$\Phi < {(T / E_C)}^{1/2}\Phi_0$. As we will show in the appendix,
moving to the pure CE adds another term, of order
$|\chi_{{}_\mathrm{L}}|(k_F \ell) (\Delta / T)^2$, which is smaller
than the difference between the GCE and the EGCE, but may still be
larger than the Landau susceptibility. All this applies to the
average susceptibility; the r.m.s. susceptibility is always assumed
to be ensemble-independent, at least to the leading order.

In the past few years it was recognized that the disorder-induced
susceptibility can be important even in the ballistic regime
$\ell > L$, and will cease to exists only in the clean limit,
$\ell > (k_F L)L$, where the disorder causes only a small perturbation
to the energy levels. However, the situation in this case is still
unclear \cite{gefen94, mccann99}, and will not be discussed in the
following.

It should be noted that all the quoted results in the diffusive regime
were obtained using diagrammatic perturbation theory, and are
therefore valid only for temperatures
above the mean level spacing $\Delta$. It was conjectured, using
different semi-quantitative arguments, that there is almost no
temperature dependence for $T < \Delta$, and that in this regime the
values of the average and typical susceptibility are given by the above
expressions evaluated at $T \sim \Delta$. \cite{serota01,serota02}

In clean mesoscopic systems the susceptibility is enhanced by boundary
effects \cite{dutch91,richter96,shapiro97,jalabert99}. Here, the
spatially limited electron motion results in oscillations in the
susceptibility as a function of the
electron density. The oscillation's period is of order $k_F L$, where
$k_F$ is the Fermi wave-vector. These oscillations are accompanied by
large paramagnetic peaks as a function of electron density. These peaks
result from the occurrence of degenerate or nearly-degenerate levels,
and are eliminated as the temperature $T$ increases above the
mean level spacing $\Delta$. The oscillations, however, persist up to
much higher temperatures, of the order of the inverse time of flight
across the sample, $\hbar v_F / L$, beyond which they decay
exponentially with the temperature ($v_F$ is the Fermi velocity). In
all the cases the susceptibility is much larger than the Landau value,
and is of order $|\chi_{{}_\mathrm{L}}|{(k_F L)}^{\alpha}$, where the
exponent $\alpha$ is of order unity, depending on whether the system is
classically integrable or chaotic, and also on the specific geometry
in the former case. Introducing disorder into the system makes these
effects decay as $\exp(-L/\ell)$, i.e., they should disappear in the
diffusive regime $\ell < L$.

Since in reality the system's size and shape cannot be set accurately,
one usually averages the results over a distribution of sizes around
an average value. This size averaging smears out the oscillations, and
reduces the size averaged GCE susceptibility to the Landau value for
temperatures higher than the mean level spacing. The size averaged
EGCE susceptibility, as well as the size averaged r.m.s.
susceptibility remain enhanced by factors which are again powers of
$k_F L$. \cite{richter96,jalabert99}
In this paper we will use the clean case only for comparison
with the diffusive case, so we will not perform size-averaging.

The paper is organized as follows: In section~\ref{sec:ii}  we will
first show that the source of difficulties in evaluating the response
of diffusive systems in the range $T < \Delta$ is related to the
problem of keeping the gauge-invariance of the results in the disorder
averaging, and that this makes Random Matrix Theory (RMT) inapplicable
for evaluating the susceptibility. This discussion will also serve to
introduce notations and formulas for usage in subsequent calculations.
In section~\ref{sec:iii} we will then examine our numerical results
for both clean and diffusive systems in the various statistical
ensembles. These will be seen to generally agree with the theoretical
predictions for $T > \Delta$, although not in all the details. For
$T < \Delta$ an unexpected strong temperature and statistical ensemble
dependence of the typical susceptibility in the diffusive regime is
observed, together with similar but weaker effects in the average
susceptibility. This will be shown in section~\ref{sec:iv} to stem from
long tails in the susceptibility distribution. It will enable us to
work out the parametric form of the r.m.s. susceptibility temperature
dependence, as well as the reasons for its statistical ensemble
dependence. Finally, we will consider interaction effects in
section~\ref{sec:v}. We will give a new way of interpreting the reasons
for interaction-induced susceptibility, as well as a numerical
demonstration of its dependence on applied in-plane magnetic field.

\section{Consequences of Gauge Invariance} \label{sec:ii}

We consider a system of spinless non-interacting electrons confined
to a finite area and moving in a random potential of impurities and a
magnetic field. The system is described by the following Hamiltonian:
\begin{eqnarray} \label{eqn:continuous}
{\hat H}= \frac{1}{2m}
{ \left( \mathbf{{\hat p}} - \frac{e}{c}\mathbf{A} \right) }^2
+V_{ \mathrm{conf} }+V_{ \mathrm{dis} },
\end{eqnarray}
where $\mathbf{{\hat p}}$ is the momentum operator, $\mathbf{A}$ is the
vector-potential, $V_{ \mathrm{conf}}$ is the confining potential and
$V_{ \mathrm{dis} }$ is the random impurity potential.

If the magnetic field is weak, its influence can be taken as a
perturbation.  Since the eigenstates in zero field can be chosen real
and are non-degenerate (due to level repulsion), the first order
correction to the energy (containing the term linear in the magnetic
field in the Hamiltonian to first order in perturbation theory)
identically vanish, i.e., there is no spontaneous magnetic moment.
The second order correction to the energy determines the susceptibility,
which is composed of the familiar Larmor diamagnetic term (containing
the term quadratic in the magnetic field in the Hamiltonian to first
order in perturbation theory) and van-Vleck paramagnetic term
(containing the term linear in the magnetic field in the Hamiltonian to
second order in perturbation theory). At finite temperature they take
the following form :
\begin{eqnarray} \label{eqn:larmor}
\chi_{{}_\mathrm{Lar}}
&=&-\frac{1}{SH^2}\frac{e^2}{mc^2}\sum_{k}f(\epsilon_k)
\langle k\left|\mathbf{A}^2\right|k\rangle , \\ \label{eqn:vv}
\chi_{{}_\mathrm{vV}}&=&-\frac{1}{SH^2}{\left(\frac{e}{2mc}\right)}^2
\sum_{k\neq l}\frac{f(\epsilon_k)-f(\epsilon_l)}{\epsilon_k-\epsilon_l}
\nonumber \\ &&\times
{\left|\langle k\left|\mathbf{A}\cdot\mathbf{{\hat p}}+
\mathbf{\hat p}\cdot\mathbf{A}\right|l\rangle\right|}^2 ,
\end{eqnarray}
where $\left|k\rangle\right.$, $\left|l\rangle\right.$ are the
eigenstates of the Hamiltonian without magnetic field and
$f(\epsilon)$ is the average occupation of a level with energy
$\epsilon$, whose form depends on the required ensemble, and
will be given at the end of this section.

It has been noted several times in the past
\cite{dutch91,serota01,serota02} that there is a large cancellation
between these two contributions. We interpret this as a consequence
of gauge invariance. If one replaces the vector potential $\mathbf{A}$
by $\mathbf{A}+\nabla \phi$, where $\phi$ is some function of the
coordinates, the observables of the system cannot change. When
treating the magnetic field as a perturbation, this means that $\phi$
must cancel between all the terms of the same order in the perturbation
theory. Indeed, it can be shown that the changes of the Larmor and
van-Vleck susceptibilities due to the gauge transformation identically
cancel. This explains why these two seemingly separate contributions are
strongly correlated and tend to cancel each other, leaving behind only
the gauge invariant part of their sum.

In fact, the expression for the Larmor susceptibility can be brought
into a form very similar to the expression for the van-Vleck
susceptibility, thus demonstrating the deep connection between them.
The resulting formula for the total susceptibility reads (in the family
of gauges obeying $\nabla \cdot \mathbf{A}=0$) :
\begin{widetext}
\begin{eqnarray} \label{eqn:susc}
\chi=-\frac{1}{2SH^2}{\left(\frac{e\hbar}{mc}\right)}^2
\sum_{k\neq l}\frac{f(\epsilon_k)-f(\epsilon_l)}{\epsilon_k-\epsilon_l}
\int_S\,d\mathbf{r}\int_S\,d\mathbf{r^{\prime}} \sum_{\alpha,\beta} &
(A_{\alpha}(\mathbf{r})-A_{\alpha}(\mathbf{r^{\prime}}))
(A_{\beta}(\mathbf{r})-A_{\beta}(\mathbf{r^{\prime}}))
\nonumber \\& \times
\psi_k(\mathbf{r}) {\nabla}_{\alpha} \psi_l(\mathbf{r})
\psi_l(\mathbf{r^{\prime}}) {\nabla^{\prime}}_{\beta}
\psi_k(\mathbf{r^{\prime}}),
\end{eqnarray}
\end{widetext}
where the indices $\alpha,\beta$ run over the two Cartesian
coordinates and the integrations are over the system's area. Since
the wave functions in this expression are most strongly correlated
when $\mathbf{r}=\mathbf{r^{\prime}}$, but the
integrand vanishes there identically, we can understand why the
resulting average susceptibility is much smaller than the Larmor or
van-Vleck terms separately.

Another lesson we can learn from this is that when trying to calculate
the disorder averaged susceptibility, one should check this does not
destroy the gauge invariance of the results. For example, the proof of
gauge invariance in our case relies on the relation
$\langle k \left| \mathbf{\hat p}/m \right| l \rangle = i(\epsilon_k-
\epsilon_l)/\hbar \langle k \left| \mathbf{\hat r} \right| l \rangle$.
This implies a connection between matrix elements of the eigenstates
and their eigenvalue difference which is usually absent in many
averaging schemes, especially RMT.

This problem does not arise in diagrammatic calculations, where the
average or r.m.s susceptibilities are expressed in terms of the energy
shifts of the diffuson and Cooperon propagators, since the equations
governing them are manifestly gauge invariant. Neither is there
a problem in the semiclassical approach, where the influence of weak
magnetic fields is taken only through the phase accumulated in closed
orbits, which depends only on the total flux through the orbits and
is thus again gauge invariant. However, for temperatures lower than
the mean level spacing those approaches are inapplicable. In this
regime RMT is usually used, but, as mentioned above,
it fails to give gauge invariant results.

To conclude this section we give the explicit form of the mean level
occupation $f(\epsilon_n)$ of the n-th single particle level. In the
GCE or EGCE it is simply the Fermi-Dirac distribution function,
\begin{eqnarray} \label{eqn:gce}
f_{\mathrm{GCE}}(\epsilon)=\frac{1}{e^{(\epsilon-\mu)/T}+1},
\end{eqnarray}
where the chemical potential $\mu$ is system independent in the GCE.
In the EGCE it is system specific, determined by the
requirement that the thermally averaged particle number in the EGCE,
$\sum_n f(\epsilon_n)$, will be equal to the constant particle number
in the sample. For the CE the occupation function can be
represented by Darwin-Fowler integrals, which are convenient for both
analytical and numerical calculations:
\begin{eqnarray} \label{eqn:ce}
f_{\mathrm{CE}}(\epsilon)=
\frac{ \mbox{\large $ \oint f_{\mathrm{GCE}}(\epsilon; z)
e^{-\Omega(z)/T} \frac{\mathrm{d}z}{z^{N+1}} $} }
{ \mbox{\large $ \oint e^{-\Omega(z)/T}
\frac{\mathrm{d}z}{z^{N+1}} $} },
\end{eqnarray}
where
\begin{eqnarray*}
\Omega(z)= -T \sum_n \mathrm{ln}
\left( 1 + ze^{ -\epsilon_n/T }\right)
\end{eqnarray*}
is the grand canonical potential, $z=\exp(\mu/T)$ is the
fugacity, and the integrations are taken around any closed contour
in the complex $z$ plane encircling the origin. This expression can
be shown to reduce to the Fermi-Dirac distribution function both for
zero temperature and for temperatures much higher than the mean level
spacing, where the integral can be evaluated using saddle-point
approximation (The saddle-point is the fugacity corresponding to the
sample-specific chemical potential in the EGCE).

\section{Numerical Simulations} \label{sec:iii}

In this section we will present the results of our numerical
simulations for the susceptibility. For this purpose we use the
familiar Anderson tight-binding Hamiltonian:
\begin{eqnarray} \label{eqn:anderson}
{\hat H} =
\displaystyle \sum_{s}\epsilon_{s} {\hat n}_{s}
- t \sum_{<s,s^{\prime}>} e^{i\theta_ {s,s^{\prime}}}
{\hat a}^{\dagger}_{s}{\hat a}_{s^{\prime}}
\end {eqnarray}
where ${\hat a}^{\dagger}_{s}$, ${\hat a}_{s}$ and ${\hat n}_{s}$
denote electron creation, annihilation and number
operators, respectively, for a state on site $s$ of a square lattice.
The first term is a random on-site potential, where $\epsilon_s$ is
chosen randomly in the range $[-W/2,W/2]$; the second is the hopping
or kinetic term, where the sum is over nearest-neighbor sites $s$
and $s^{\prime}$, $t$ is the overlap integral, and the phase
$\theta_{s,s^{\prime}}=e/\hbar c \int^{s}_{s^{\prime}}\mathbf{A}
\cdot\mathrm{d}\mathbf{r}$ gives the influence of the external
magnetic field. The Anderson model has the advantage of being
discrete, but, unlike numerical discretization of a continuous
Hamiltonian, it has all the required properties of a Hamiltonian
(i.e., being Hermitian, gauge-invariant, etc.)

In the calculation we use expressions for the Larmor and van-Vleck
susceptibilities for the Anderson Hamiltonian. Since the magnetic
field is assumed small, we first expand the phase exponent containing
the magnetic field in Eq.~(\ref{eqn:anderson}) in series. As for the
continuous Hamiltonian (\ref{eqn:continuous}), the linear term in the
field taken to first order in perturbation theory identically vanishes.
We are left with the second order corrections to the energy -- the
quadratic term in the field taken to first order in perturbation
theory (Larmor susceptibility), and the linear term in the field taken
to second order in perturbation theory (van-Vleck susceptibility).We
thus get formulas analogous to Eqs.~(\ref{eqn:larmor}),(\ref{eqn:vv})
for the susceptibility, which are used in all the subsequent
calculations. The level occupations are calculated in the required
ensembles, using Eqs.~(\ref{eqn:gce}),(\ref{eqn:ce}). We note that
even though the contour of integrations in
Eq.~(\ref{eqn:ce}) can be evaluated in principle using any path in the
complex plane encircling the origin, to get sensible results in
numerical integration one should use a contour passing through or near
the sample specific saddle point (the EGCE fugacity) in a direction
where this point is a maximum of the integrand. A suitable form of the
contour is thus a circle around the origin, and this choice was used
in our calculations.

The calculations were usually made (unless
otherwise specified) on a $17\times 24$ lattice (the sizes were chosen
mutually prime so that there's no degeneracy even in the clean limit,
and our non-degenerate perturbation theory is applicable).
We have made calculations both on clean ($W=0$) systems and disordered
systems in the diffusive regime, especially with the disorder values
$W=2.0$ and $W=4.0$. For the diffusive systems the results were averaged
over 2500 realizations of disorder, unless otherwise specified.
Dependence on the system's size was deduced from comparison to the
results on $13\times 19$ and $8\times 13$ lattices. The susceptibility
is usually plotted as a function of electron filling (i.e., the ratio
of the number of electrons and the number of lattice sites). To enable
comparison this is done not only in the CE or EGCE, but also in the
GCE, in which case we refer to the ensemble and thermally averaged
filling.

\subsection{Macroscopic Clean Systems}

For comparison with subsequent results for finite systems, we first
show the orbital susceptibility of the tight-binding Hamiltonian
(\ref{eqn:anderson}) for infinite clean lattice. The zero temperature
susceptibility was calculated in Ref.~\onlinecite{skudlarski91}, and
is given by the following expression :
\begin{eqnarray}
\chi(T=0,\epsilon_F) = -\frac{a^2e^2}{24\hbar^2c^2}
\left[ \left(\epsilon_F^2-8t^2 \right)\rho(\epsilon_F) -E(\epsilon_F)
\right],
\end{eqnarray}
where
\begin{eqnarray*}
\rho(\epsilon)=\frac{1}{2\pi^2t}
K \left( \sqrt{1-\left(\epsilon/4t\right)^2} \right)
\end{eqnarray*}
is the density of states ($K$ denotes the complete elliptic integral
of the second kind), $E(\epsilon_F)$ is the total energy of the filled
states, and $a$ is the lattice constant. At the band edges we get the
value $-a^2e^2t/12\pi\hbar^2c^2$, which is the Landau susceptibility
with the appropriate
effective mass. At nonzero temperatures the susceptibility can be
evaluated from its zero temperature value using the connection
$\chi(T,\mu)=\int \chi(0,\epsilon) f^{\prime}_{\mathrm{GCE}}
(T,\epsilon) \mathrm{d} \epsilon$, where $f^{\prime}_{\mathrm{GCE}}$
is the energy derivative of the Fermi-Dirac distribution function (of
course, in this macroscopic limit the difference between statistical
ensembles disappears).

\begin{figure}
\includegraphics[width=8.5cm,height=9cm]{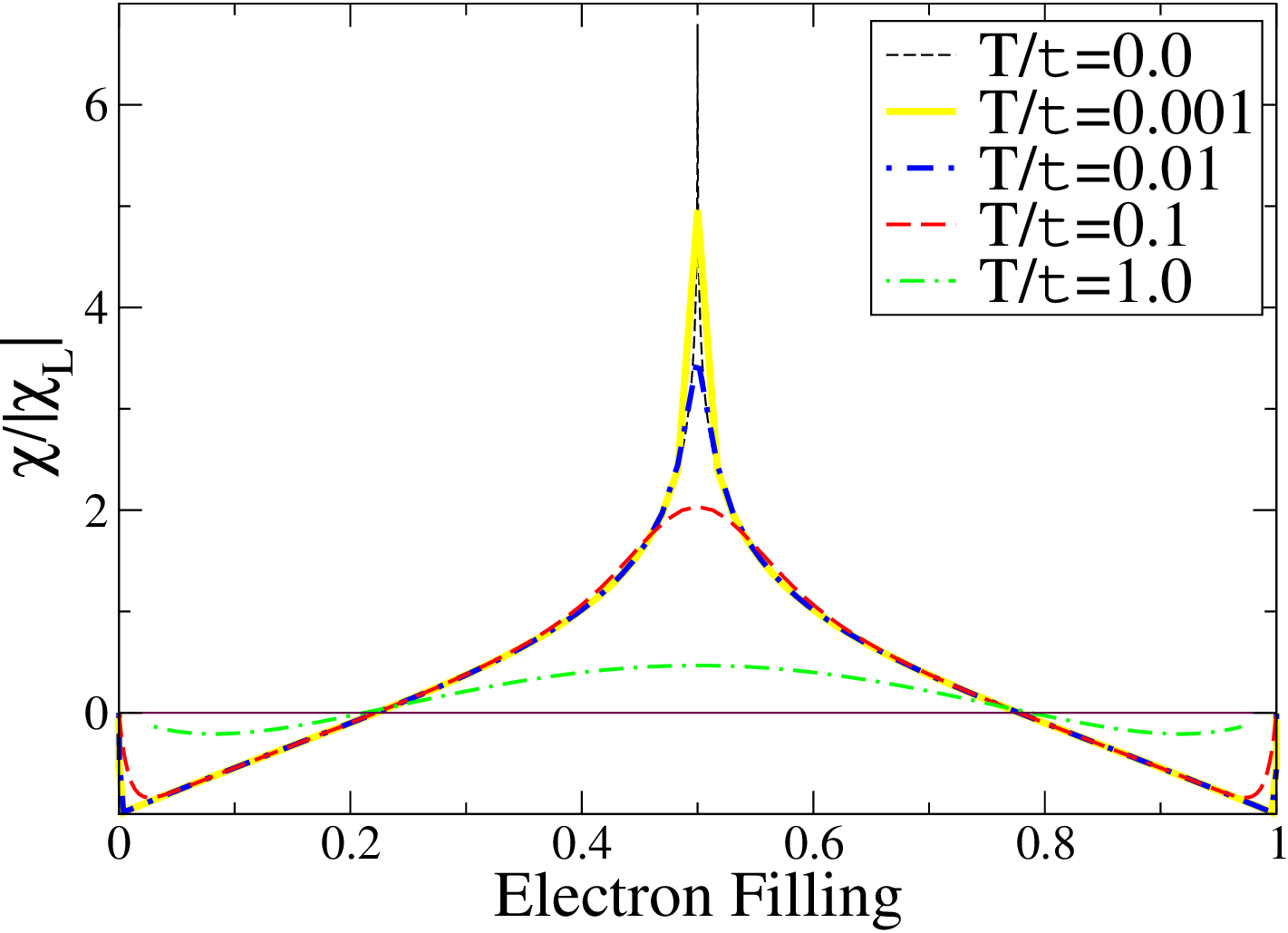}
\caption{\label{fig:macroc}The orbital susceptibility of an infinite
ordered lattice as a function of filling for various temperatures.}
\end{figure}

The susceptibility is plotted in Fig.~\ref{fig:macroc} as a function
of the filling
$\nu=\int \rho(\epsilon) f_{GCE}(\epsilon) \mathrm{d}\epsilon$. We
can see that the susceptibility has a diamagnetic Landau value only at
the band edges at zero temperature, and that its sign changes to
paramagnetic near the band center. At half-filling it actually exhibits
a logarithmic singularity at zero temperature. Those non-analytic
behaviors at the band's center and edges cause noticeable temperature
dependence of the susceptibility there, even for temperatures much lower
than the Fermi energy (the only energy scale in macroscopic clean 
systems),
although in the rest of the band temperature has a significant influence
only when it reaches the Fermi energy. This behavior is connected to the
inclusion of the magnetic field only as a phase factor in the hopping
amplitude. However, if we remember in the sequel to compare our results
for finite, possibly disordered systems to this macroscopic limit, we
would not have any trouble interpreting the results.

\subsection{Mesoscopic Clean Systems}

\begin{figure}
\includegraphics[width=8.5cm,height=!]{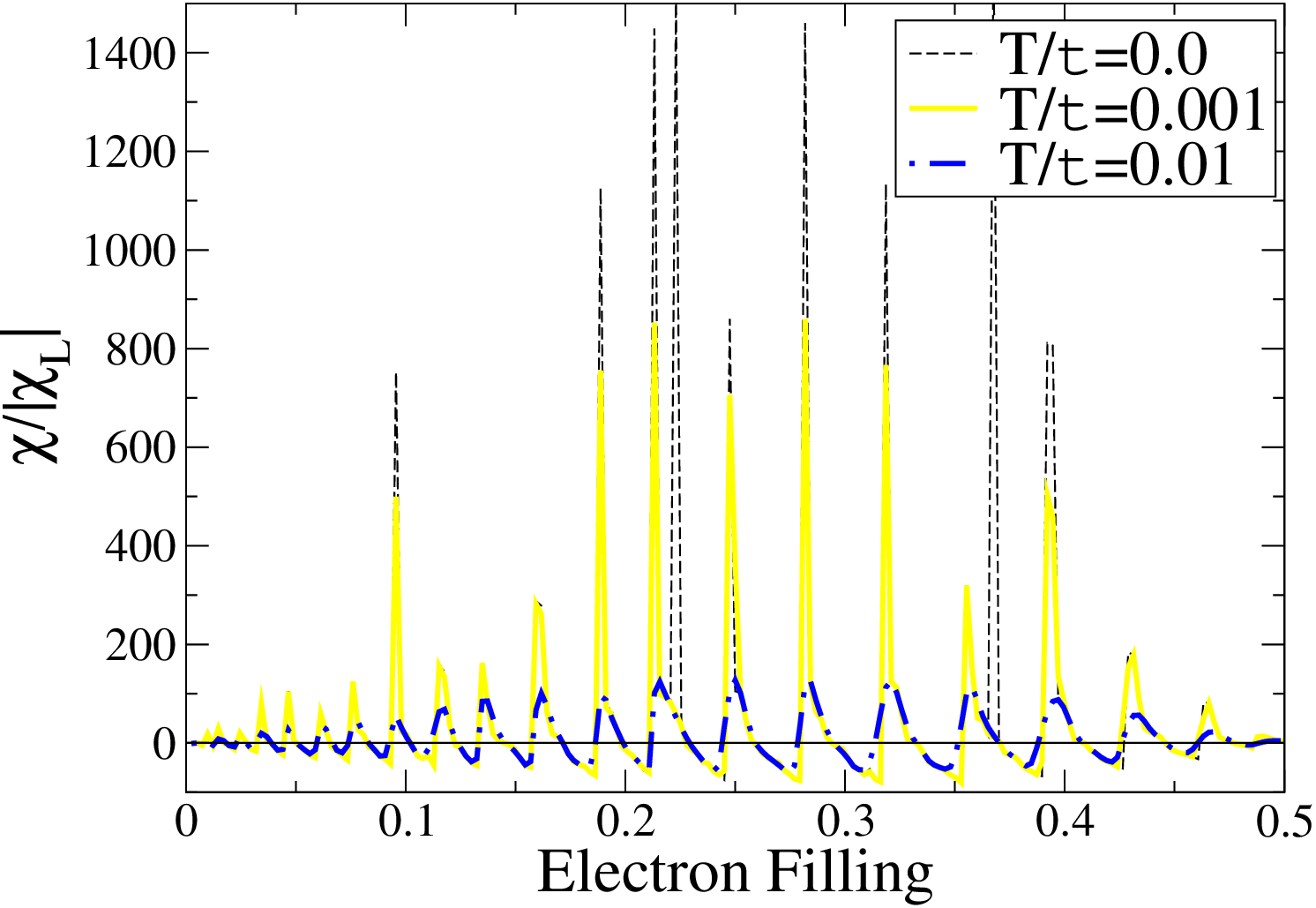}
\vskip 1.0cm
\includegraphics[width=8.5cm,height=!]{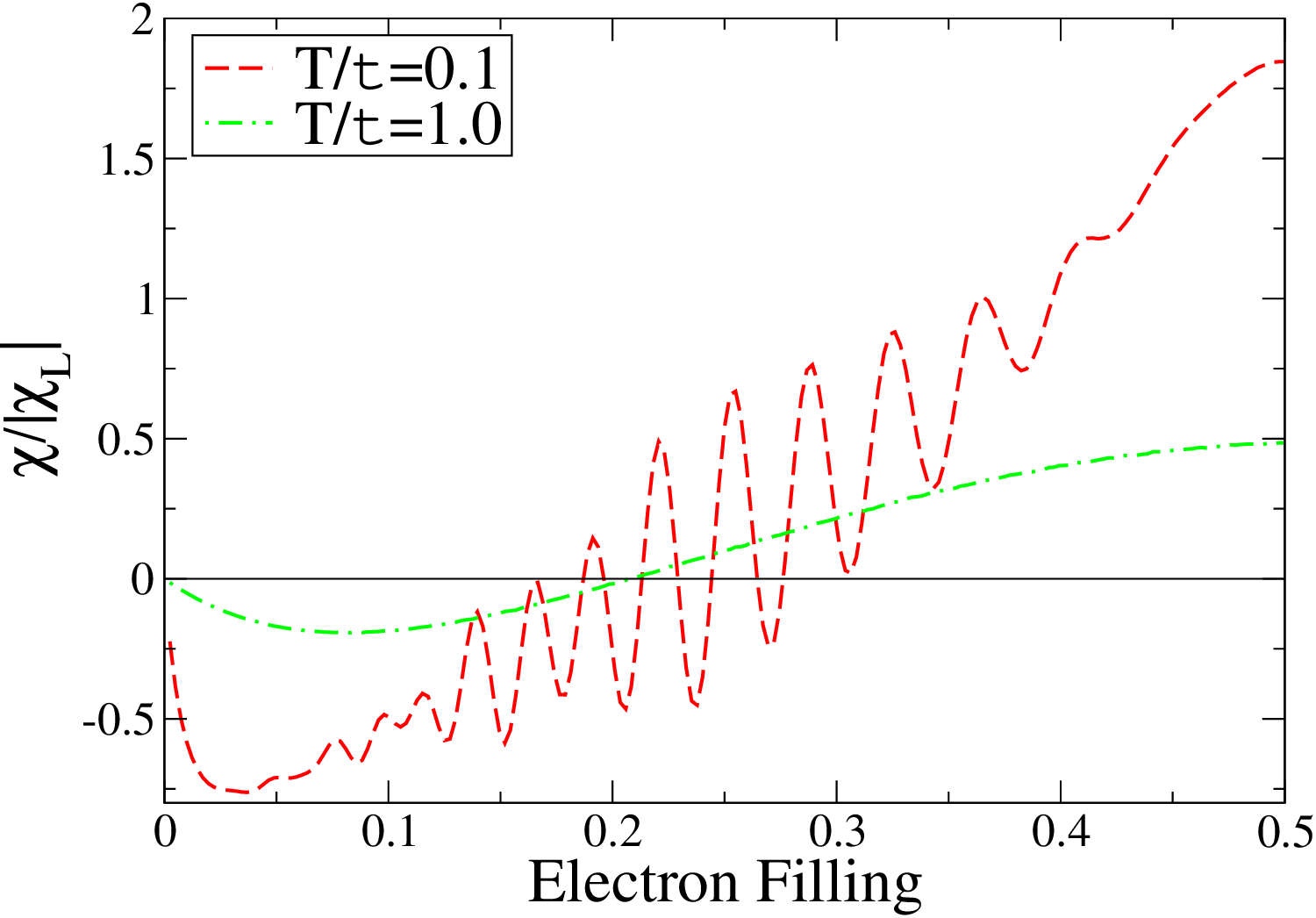}
\caption{\label{fig:mesocegc} The orbital susceptibility of a finite
clean lattice as a function of filling for various temperatures in
the EGCE. Low temperature results are presented in the upper panel,
high temperature results are presented in the lower panel.}
\end{figure}

We now turn to finite clean systems. The results for the susceptibility
of a
system on a $17\times 24$ lattice are shown in Fig.~\ref{fig:mesocegc}
in the EGCE (which is equivalent to the GCE in the absence of disorder),
for various temperatures. The mean level spacing $\Delta$ is
approximately $0.02t$. We can see here all the expected phenomena
described in the introduction. At zero temperature, or finite
temperatures much smaller than the mean level spacing, there are very
large paramagnetic peaks, occurring where the level spacing is small.
Their exact position and height strongly depend on the system's area
and the ratio of its length and width. Those peaks are smeared out
when the temperature approaches the level spacing, leaving behind
oscillations whose amplitude and frequency scale roughly as $L^{3/2}$
and $L^{-1}$, respectively, as in continuous square geometries
\cite{shapiro97}.
They still have a paramagnetic tendency, although much smaller than
the low temperature peaks. These oscillations are smallest near the
band edges or center, and are maximal near quarter filling. This is in
contrast with the macroscopic susceptibility, which is maximal near
the band center. The oscillations disappear, and the susceptibility
assumes its macroscopic value (including the subsequent temperature
dependence) for temperature around $0.2t\approx 10\Delta$, which can
be attributed to the inverse time of flight across the sample
$\hbar v_F/L$ for all but the lowest fillings.

\begin{figure}
\includegraphics[width=8.5cm,height=!]{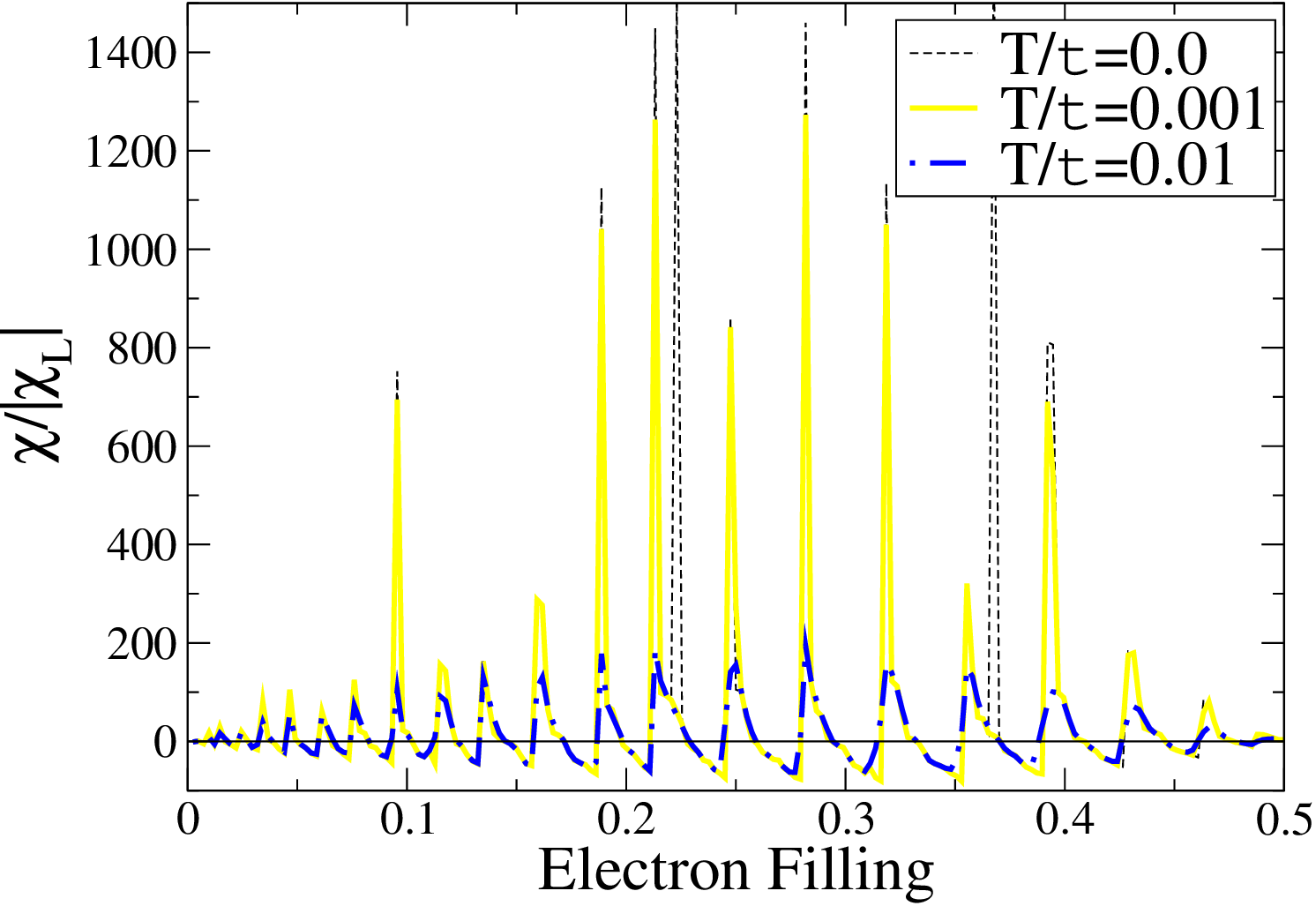}
\vskip 1.0cm
\includegraphics[width=8.5cm,height=!]{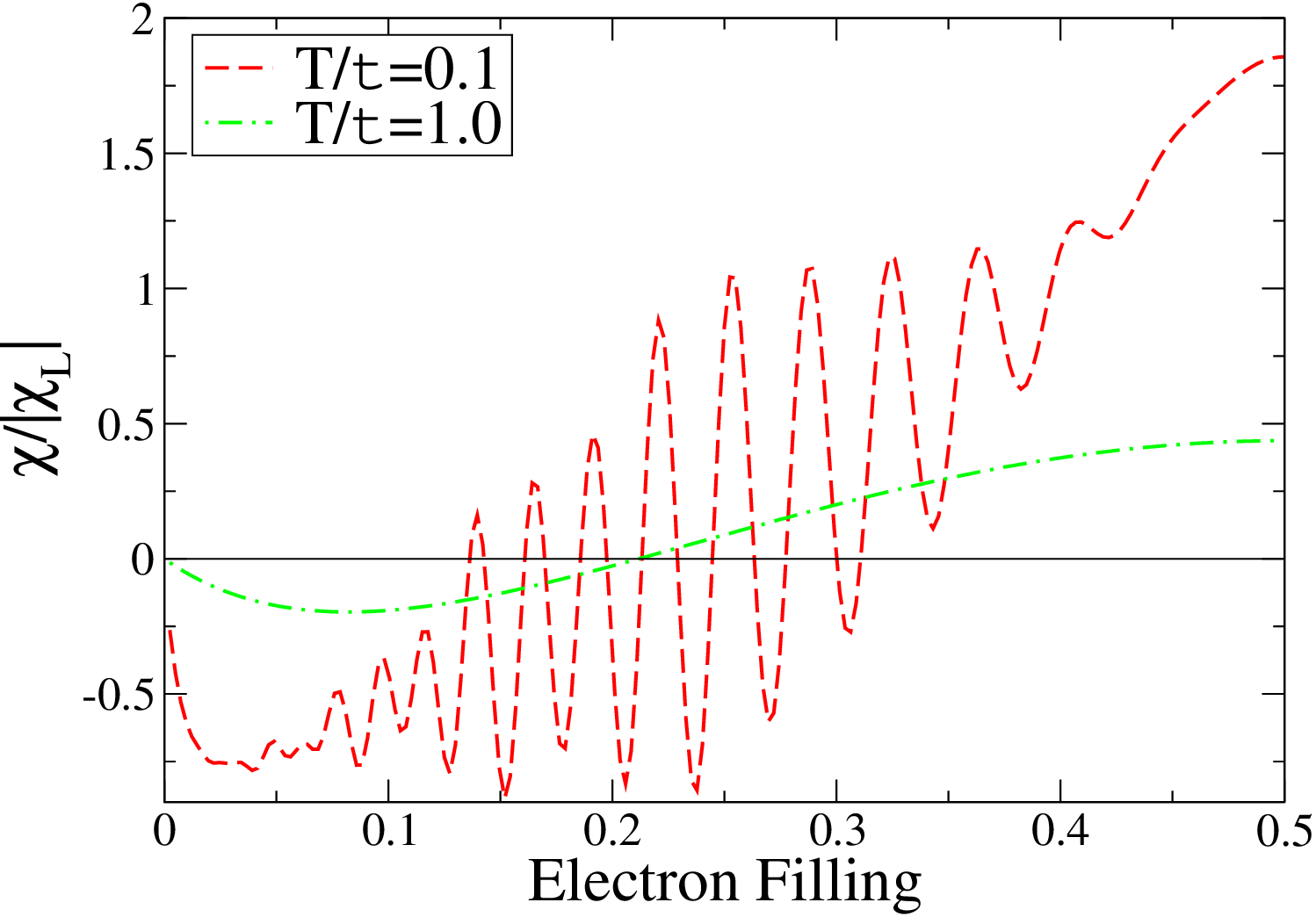}
\caption{\label{fig:mesocce} The orbital susceptibility of a finite
ordered lattice as a function of filling for various temperatures in
the CE. Low temperature results are presented in the upper panel,
high temperature results are presented in the lower panel.}
\end{figure}

As regards the CE, The results for the same values of parameters are
shown in Fig.~\ref{fig:mesocce}. We can see that moving from the EGCE
to the CE is qualitatively equivalent to lowering the temperature in
the EGCE, as noted by various authors in the past
\cite{denton73, kamenev97}. Hence, the peaks
and oscillations are somewhat larger in the CE, and thus persist to
higher temperatures. The difference disappears at both zero temperature
and very high temperatures (of the order of the temperature where the
mesoscopic effects in the EGCE disappear), but is observable for
intermediate temperatures. However, even when non-negligible, this
difference is smaller than the EGCE susceptibility.

\subsection{Mesoscopic Disordered Systems, GCE}

\begin{figure}
\includegraphics[width=8.5cm,height=!]{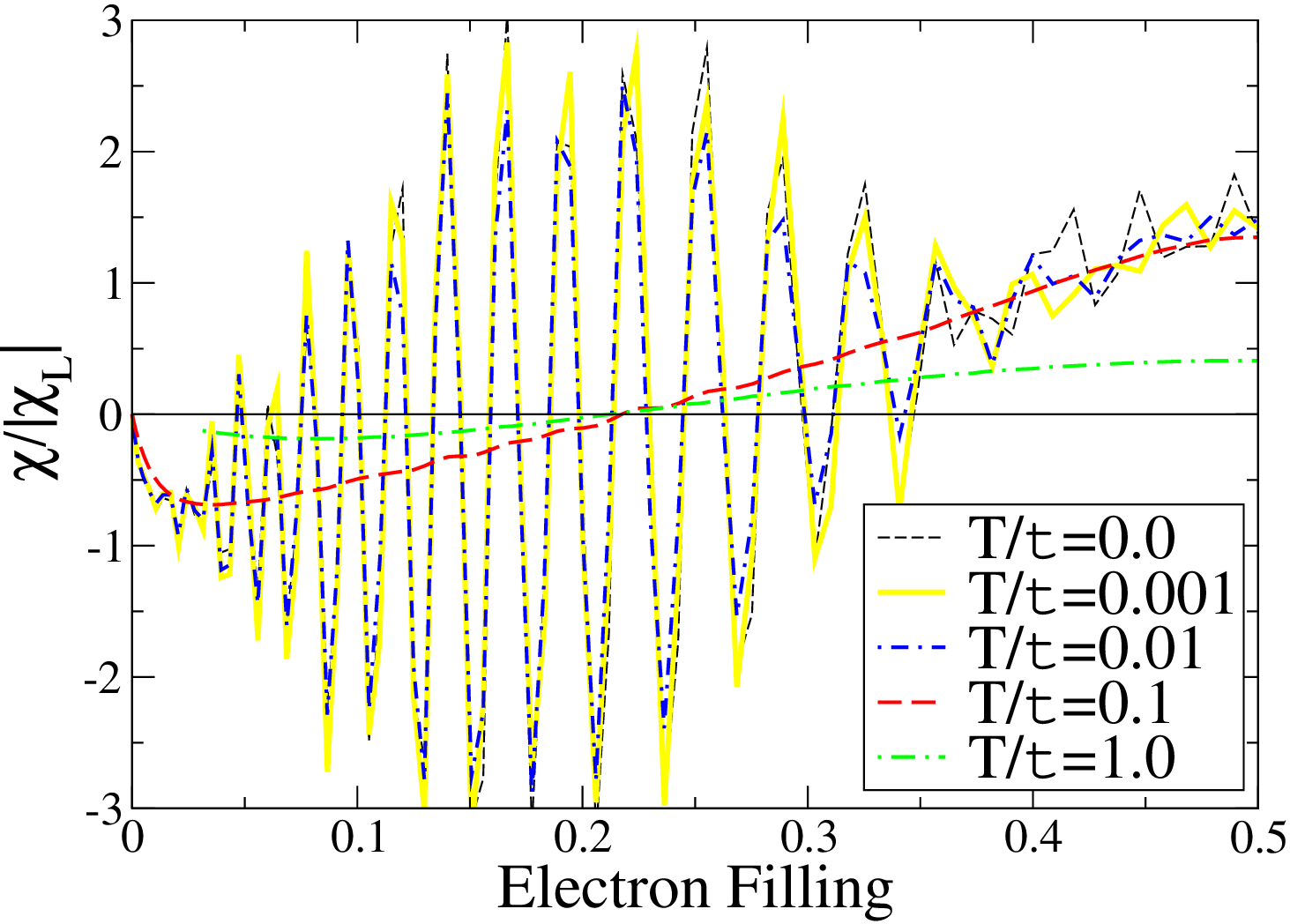}
\vskip 1.0cm
\includegraphics[width=8.5cm,height=!]{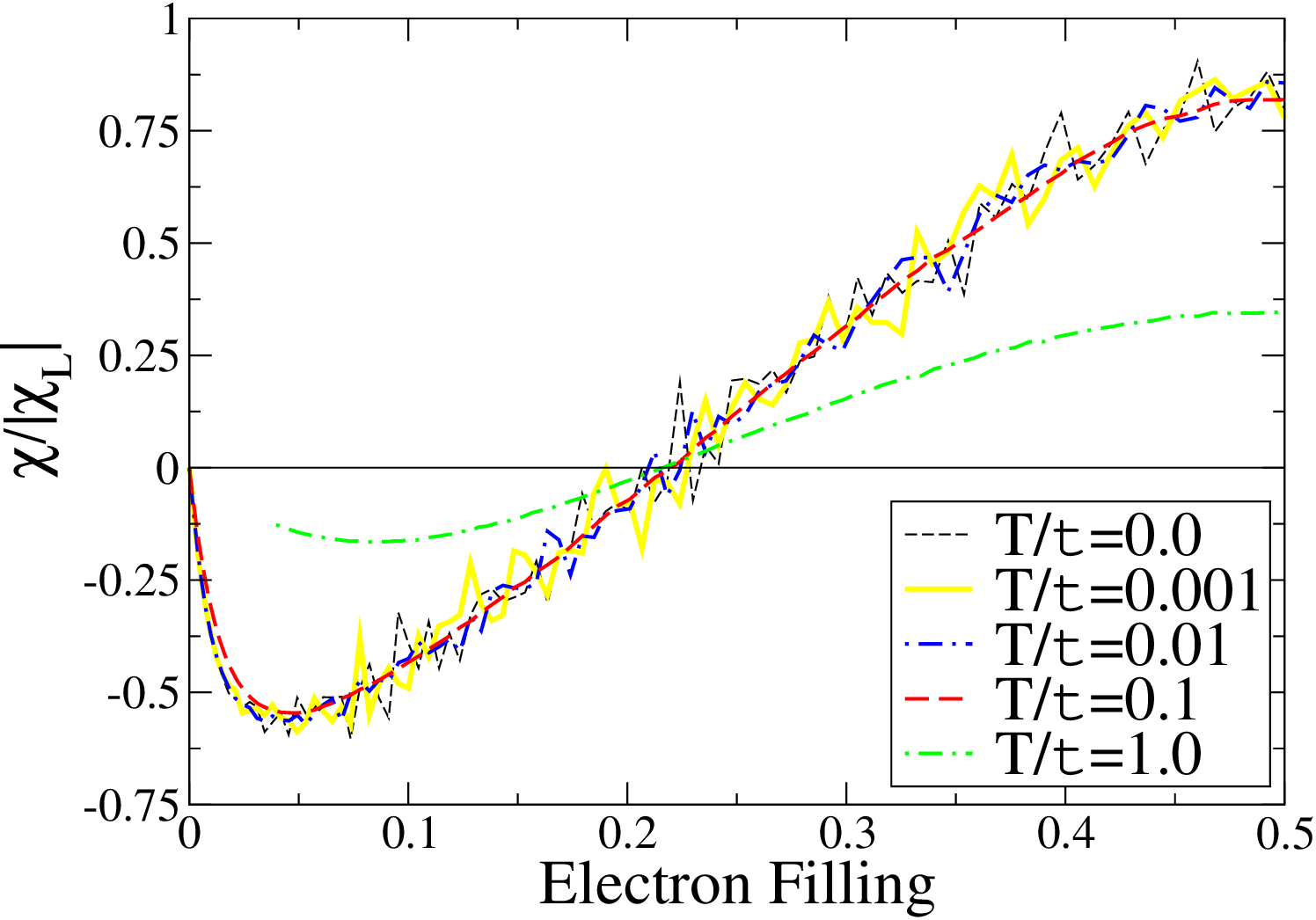}
\caption{\label{fig:mesodgc} The disorder averaged orbital
susceptibility of a finite disordered lattice as a function of
filling for various temperatures in the GCE, for both $W=2.0$ (upper
panel) and $W=4.0$ (lower panel).}
\end{figure}
We now introduce disorder into the system. The results for the
average susceptibility in the GCE are shown in
Fig.~\ref{fig:mesodgc}. The GCE susceptibility
is expected, as noted in the introduction, to be equal to the
macroscopic susceptibility, shown in Fig.~\ref{fig:macroc}.
Indeed, the general behavior of the GCE susceptibility resembles the
macroscopic susceptibility (especially in its sign as a function of
filling). This is immediately obvious regarding the case $W=4.0$, and
is also true for $W=2.0$ if the oscillations as a function of filling
are averaged out (These oscillations will be discussed below). In
addition, the GCE susceptibility (ignoring again the
oscillating part for $W=2.0$) shows no temperature dependence, up to
about $T\approx 0.1t$. This is in accordance with the macroscopic
behavior if we remember that there are no zero temperature
singularities as a function of filling for finite samples, so only the
high temperature macroscopic temperature dependence persists.

There are, however, two modifications compared to the macroscopic
results. The first is that the
susceptibility of the disordered mesoscopic systems is clearly smaller
than the macroscopic value. It also decreases when disorder increases,
although quite slowly. This can be interpreted as stemming from 
corrections to susceptibility (beyond the leading order macroscopic
value), which are expected to vary as ${(k_F \ell)}^{-1}$. Since
${(k_F \ell)}^{-1}$ is not so small in our simulations, these
corrections can be important and lead to a weak disorder dependence,
and their sign is apparently negative.

A second and more pronounced difference between the GCE average
susceptibility and its macroscopic value is the above mentioned
filling-dependent oscillations for $W=2.0$. The oscillations'
position is the same as the oscillations in the ordered systems
shown above in Fig.~\ref{fig:mesocegc}, i.e., of periodicity varying
as $L^{-1}$. The amplitude of the oscillations, however, shows little
dependence on the system's size, but decreases strongly when disorder
or temperature increase. This effect is due to the fact that even
though for $W=2.0$ the systems of the size considered are regarded
diffusive, showing, for example, a Wigner level spacing distribution,
the elastic mean free path $\ell$ is only about a third of the size
of the considered systems, so that ballistic effects can still be
observed. In fact, the density of states still shows ballistic
oscillations for $W=2.0$, which are smeared out as disorder increases
toward $W=4.0$. Thus, the susceptibility at $W=2.0$ shows a mixture of
diffusive and ballistic behaviors. We note that size-averaging will
smear out these oscillations.

\begin{figure}
\includegraphics[width=8.5cm,height=!]{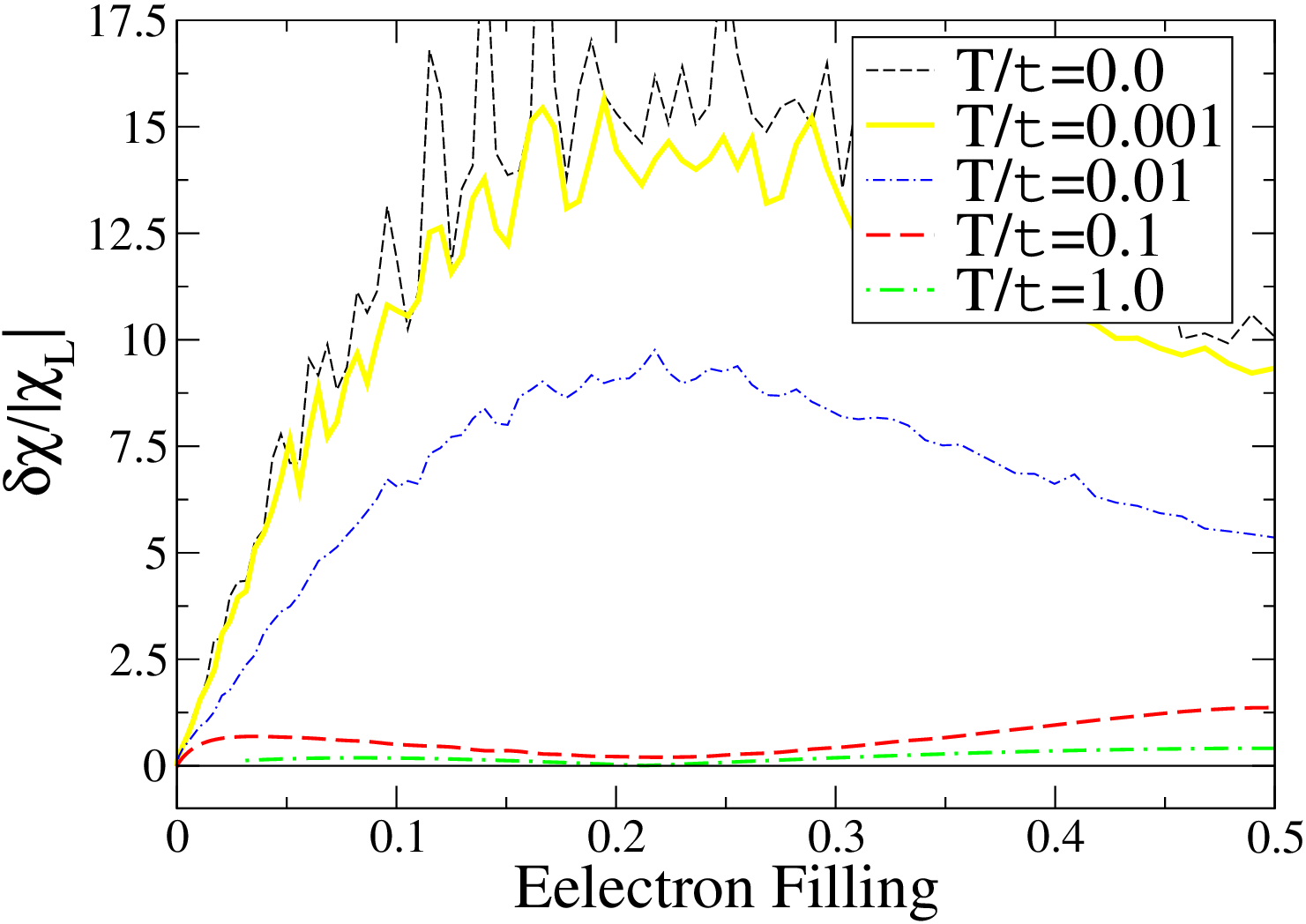}
\vskip 1.0cm
\includegraphics[width=8.5cm,height=!]{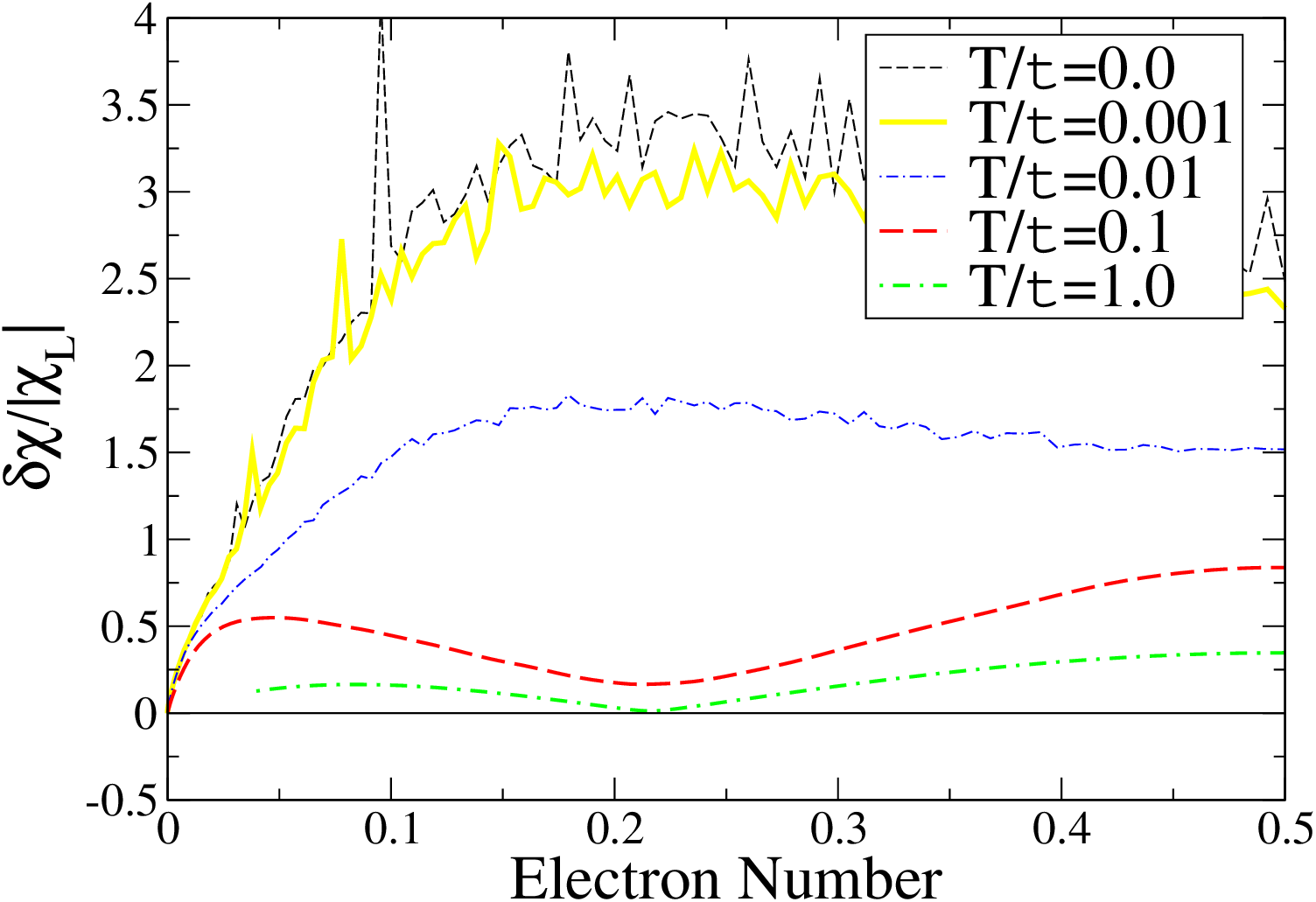}
\caption{\label{fig:mesodgc2} The disorder averaged r.m.s. orbital
susceptibility of a finite disordered lattice as a function of
filling for various temperatures in the GCE, for both $W=2.0$ (upper
panel) and $W=4.0$ (lower panel).}
\end{figure}

Referring now to fluctuations in the susceptibility, the results for
the r.m.s. susceptibility of the same system are shown in
Fig.~\ref{fig:mesodgc}. We can clearly see that the r.m.s.
susceptibility is much larger than the average. It also decreases
strongly with increasing disorder, implying the expected
$|\chi_{{}_\mathrm{L}}|(k_F \ell)$ dependence (This can be seen
quantitatively by
noting that from Fermi's golden rule one can estimate $\ell$ to vary
as $W^{-2}$, which is quite accurately the W dependence of the r.m.s
susceptibility in our data). Since $k_F \ell \sim g$, where $g$ is
the dimensionless conductance, the enhancement factor agrees well
with the values of $g \approx 20$ and $g \approx 3$ for $W=2.0$ and
$W=4.0$ respectively. (The values $g$ were calculated from 
non-universal corrections to the inverse participation ratio
\cite{prigodin98}, and agrees well with estimates based on Fermi's
golden rule.)

A surprising feature of the results is that the r.m.s. susceptibility
depends strongly on temperature even for temperatures much smaller
from the mean level spacing $\Delta$ (which is, approximately, $0.022t$
and $0.025t$ for $W=2.0$ and $W=4.0$, respectively) This will be
discussed later on. Moreover, the r.m.s. susceptibility becomes equal
to the average susceptibility for a temperature of about
$0.1t\approx 5\Delta$
for both $W$ values. As mentioned in the introduction, significant
fluctuations in the susceptibility (i.e., larger than the average
value) are expected to persist even in temperatures higher than the
Thouless energy. However, since the enhancement factor $k_F \ell$ is
not so large in our simulations as in real systems, absence of
fluctuations in temperatures above the Thouless energy is quite
reasonable. On the other hand, fluctuations below the Thouless energy
should be significant. This implies interpreting the temperature
$0.1t\approx 5\Delta$, for which the difference between the average
and r.m.s. susceptibilities disappears, as the Thouless energy
times a numerical factor (note that our estimate for $g$ indicates a
much higher Thouless energy for $W=2.0$.)

However, although this temperature goes as $L^{-2}$ as the Thouless
energy should vary, it shows very little if any disorder dependence, so
its identification with the Thouless energy is quite problematic. This
may support the claim of some authors \cite{altland95} that in the
logarithmic factor $\mathrm{ln}(E_C/T)$ in the r.m.s susceptibility,
the Thouless energy should be replaced by some lower energy scale. Our
results seem to suggest that this scale, which is disorder independent,
is connected to the mean level spacing $\Delta$.

\subsection{Mesoscopic Disordered Systems, EGCE}

\begin{figure}
\includegraphics[width=8.5cm,height=!]{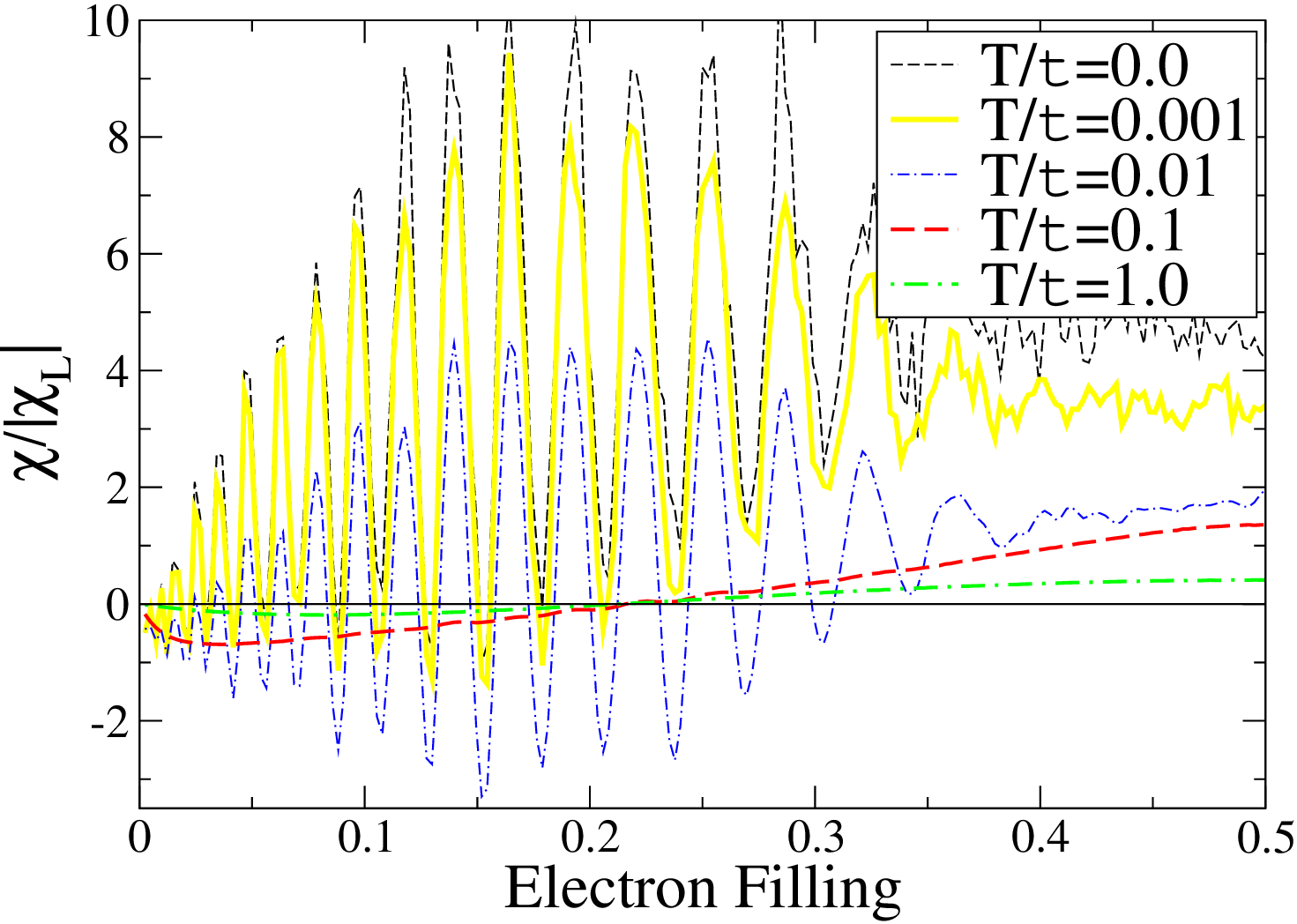}
\vskip 1.0cm
\includegraphics[width=8.5cm,height=!]{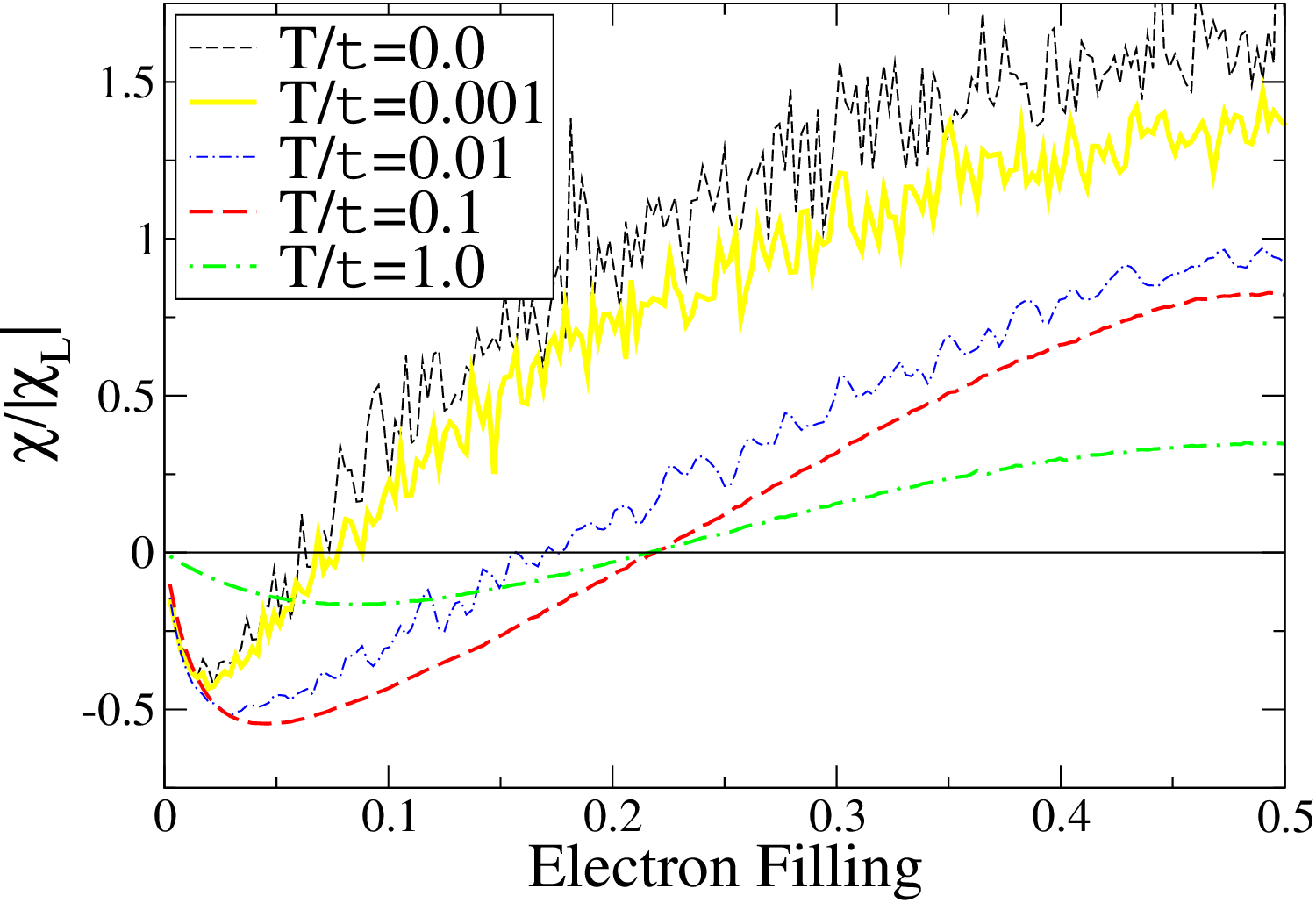}
\caption{\label{fig:mesodegc} The disorder averaged orbital
susceptibility of a finite disordered lattice as a function of
filling for various temperatures in the EGCE, for both $W=2.0$ (upper
panel) and $W=4.0$ (lower panel).}
\end{figure}

We now turn to discuss disordered systems in the EGCE. The results for
the average susceptibility are shown in Fig.~\ref{fig:mesodegc}. We
see again the oscillations in the $W=2.0$ susceptibility. Their
explanation is the same as in the GCE case discussed above, so we now
ignore them and refer to the smooth part of the susceptibility.

As is expected, the difference between the average EGCE susceptibility
and the GCE susceptibility is positive. It is also much larger
than the GCE susceptibility except near the band edges.
This is somewhat problematic since precisely at the band edges the
macroscopic susceptibility equals the Landau value. It may be explained
by the fact that those states are more localized than the other states,
and thus their susceptibility is not so much enhanced.

The difference between the average EGCE and GCE susceptibilities is
strongly disorder dependent, confirming the identification of the
zero temperature enhancement factor as $k_F \ell$ (As for the r.m.s.
GCE susceptibility discussed above, it has quite accurately a $W^{-2}$
dependence). Temperature dependence shows up for temperatures higher
than for the r.m.s. GCE susceptibility discussed above, but much lower
than the mean level spacing. In fact the difference between the EGCE
and GCE practically disappears for $T=0.02t \approx \Delta$ and
$T=0.01t \approx 0.4\Delta$ for $W=2.0$ and $W=4.0$ respectively.
Although this temperature is much smaller than the Thouless energy
(based on our estimate $g \approx 20$ and $g \approx 3$ for $W=2.0$
and $W=4.0$ respectively), its $L^{-2}$ size dependence as well
as its disorder dependence indicate that it might be considered as
connected with the Thouless energy reduced by a numerical factor. We
may note, however, that the disorder dependence of this temperature is
quite weaker than the corresponding dependence of the Thouless energy
(the above mentioned $W^{-2}$ law). Combined with the quite low value
of this temperature, its meaning is not completely clear.

\begin{figure}
\includegraphics[width=8.5cm,height=!]{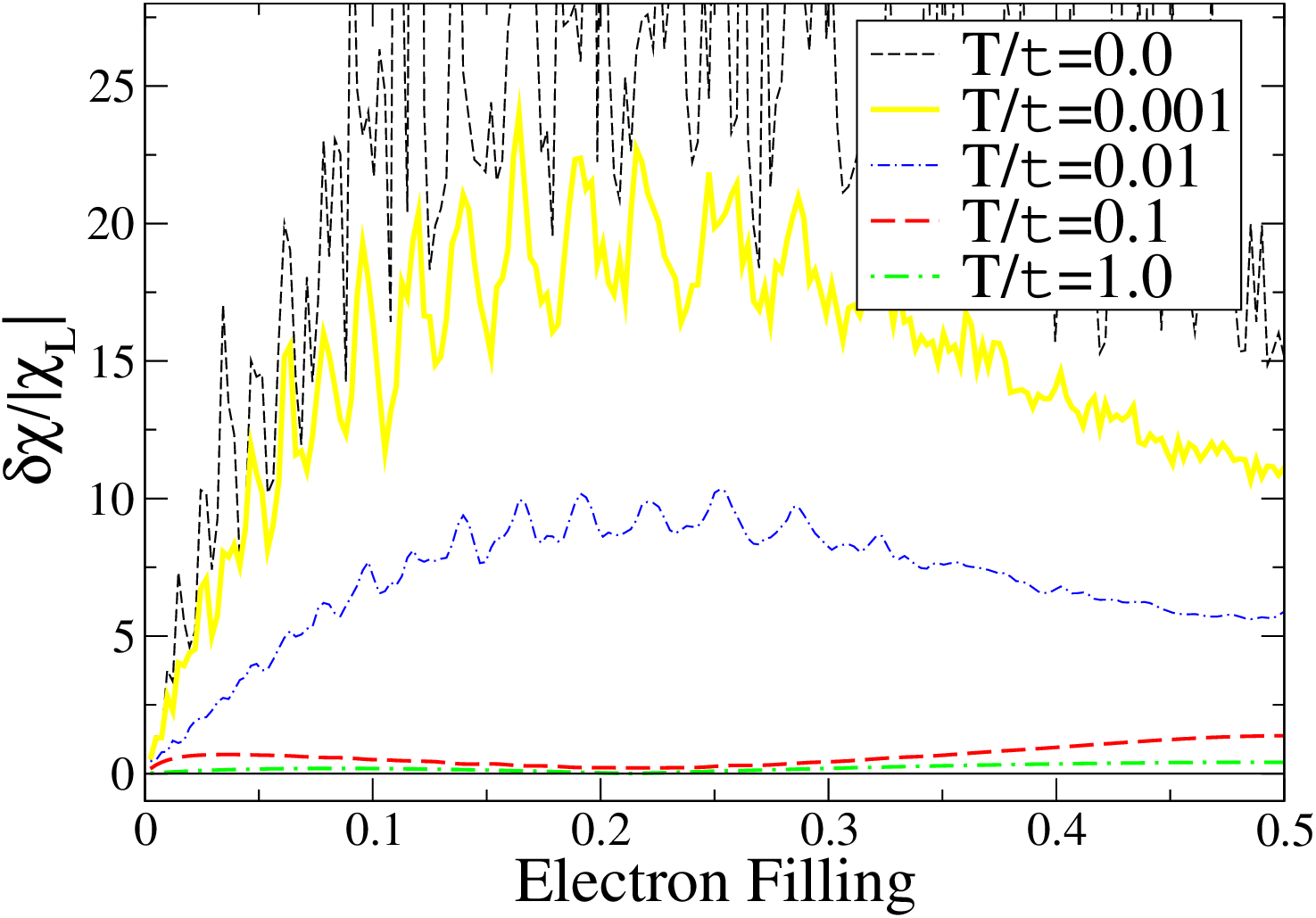}
\vskip 1.0cm
\includegraphics[width=8.5cm,height=!]{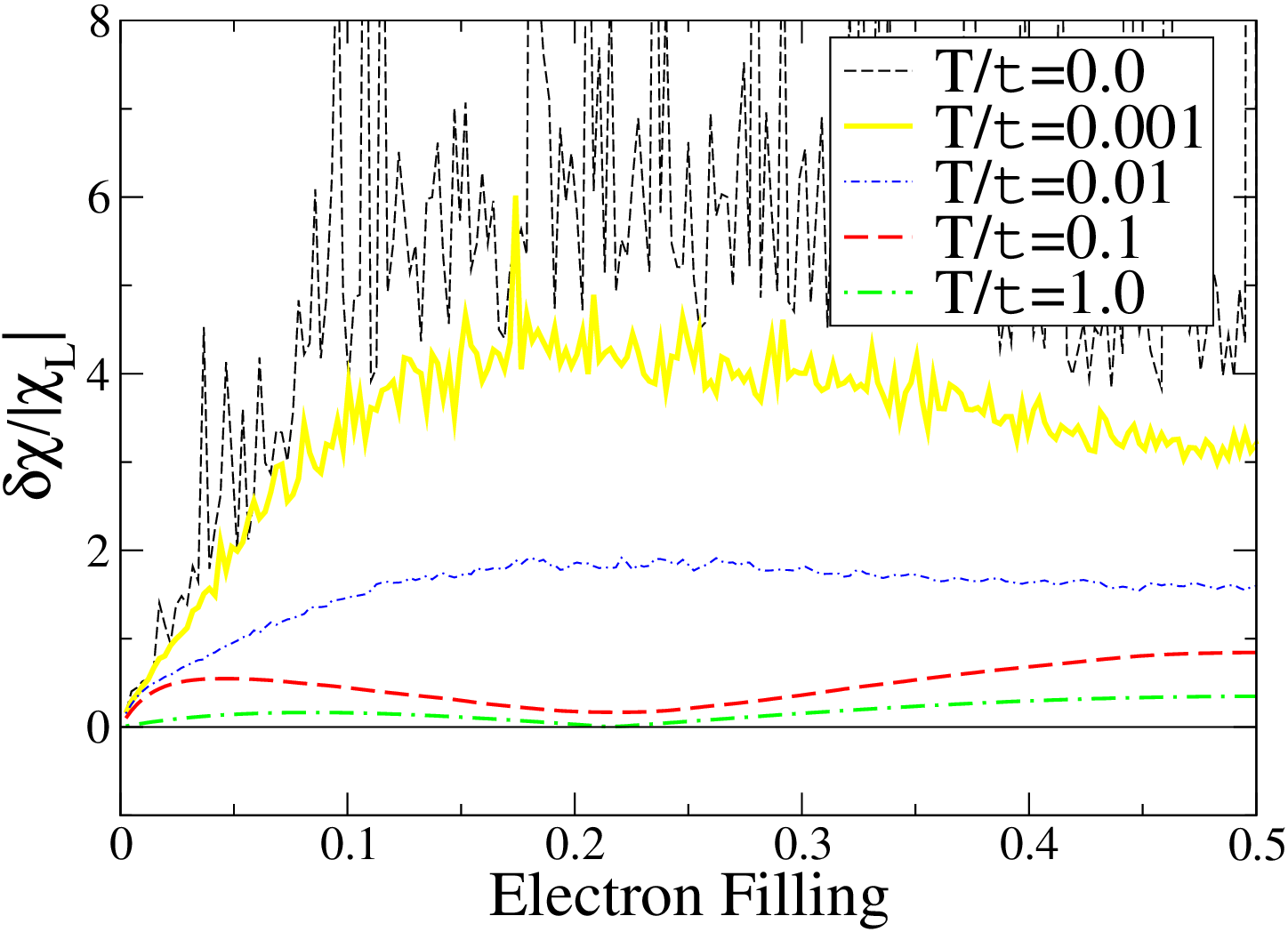}
\caption{\label{fig:mesodegc2} The disorder averaged r.m.s. orbital
susceptibility of a finite disordered lattice as a function of
filling for various temperatures in the EGCE, for both $W=2.0$ (upper
panel) and $W=4.0$ (lower panel).}
\end{figure}

As for the fluctuations in the susceptibility, The results for
the r.m.s. susceptibility are shown in Fig.~\ref{fig:mesodegc2}. As
is usually conjectured, it behaves quite similarly
to the GCE r.m.s. susceptibility, and all the discussion regarding the
GCE applies also here. The only exception is the very low temperature
regime, (below $0.01t\approx 0.5\Delta$), where the EGCE r.m.s.
susceptibility is well above the GCE values, the difference being
larger for $W=2.0$. It is also much more ``noisy'', as can be clearly
seen in the $T=0$ curves. We will discuss this feature in the next
section.

\subsection{Mesoscopic Disordered Systems, CE}

To conclude this section, we move to the CE. The results for both the
averaged and r.m.s. susceptibilities are very similar to the EGCE
results, the difference being only few percents of the EGCE values and
thus not shown. This is quite expectable for $T > \Delta$, since, as
we show in the appendix, the difference between the CE and EGCE is
parametrically smaller than the difference between the EGCE and the
GCE.

We may note, however, that the difference between the CE and EGCE is
more pronounced for the r.m.s susceptibility at very low temperatures
(below $0.01t\approx 0.5\Delta$), where there is also a difference
between the EGCE and the GCE r.m.s. susceptibilities. We believe those
two phenomena has a similar origin, and will discuss it in the next
section.

\section{Zero Temperature Susceptibility Distribution} \label{sec:iv}

In this section we will examine the zero temperature susceptibility
distribution, and use it to explain the strong dependence of the
typical and average susceptibilities on both temperature and the
statistical ensemble for temperatures lower than the mean level
spacing.

\begin{figure}
\includegraphics[width=8.5cm,height=8cm]{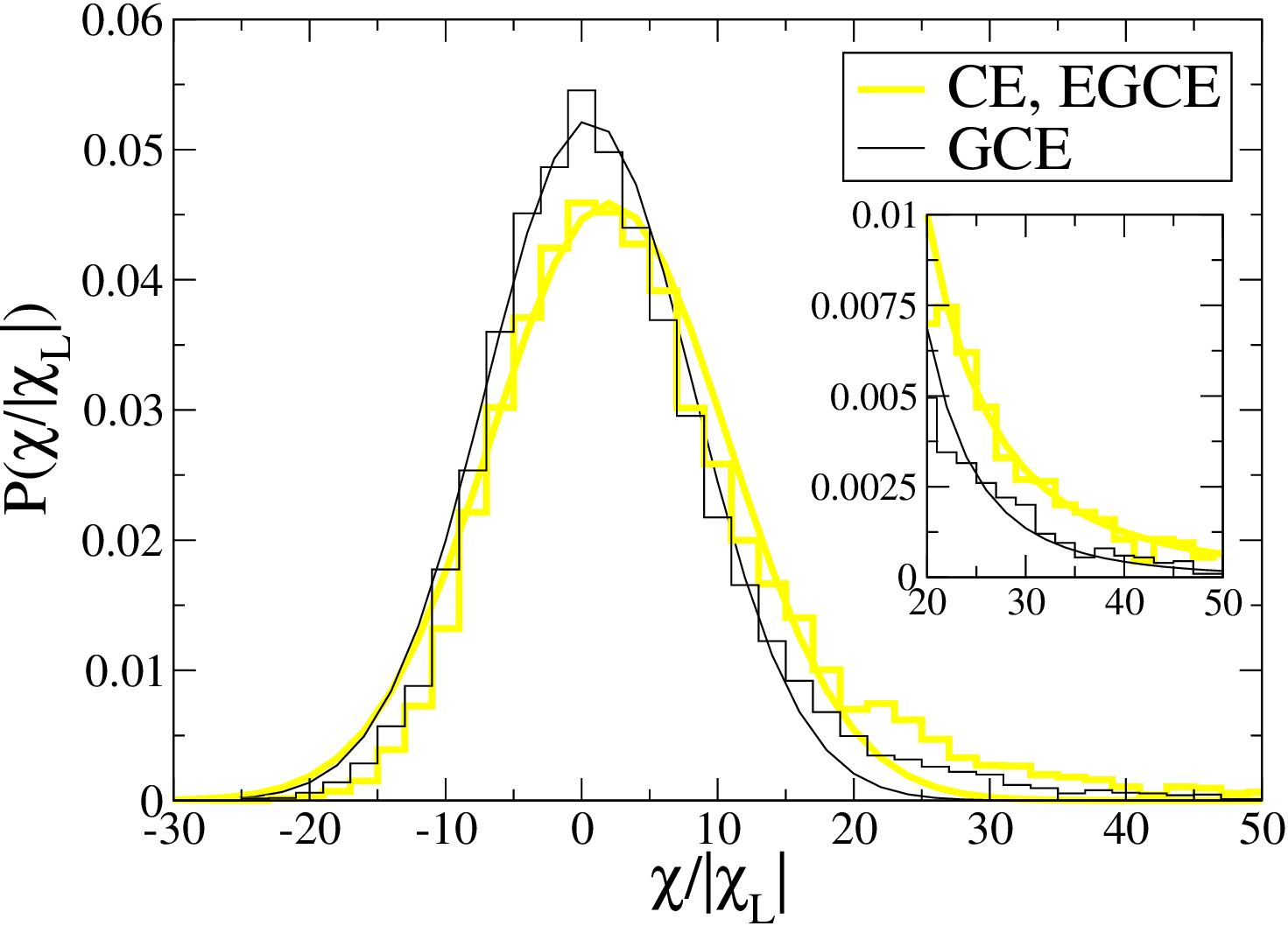}
\vskip 1.3cm
\includegraphics[width=8.5cm,height=8cm]{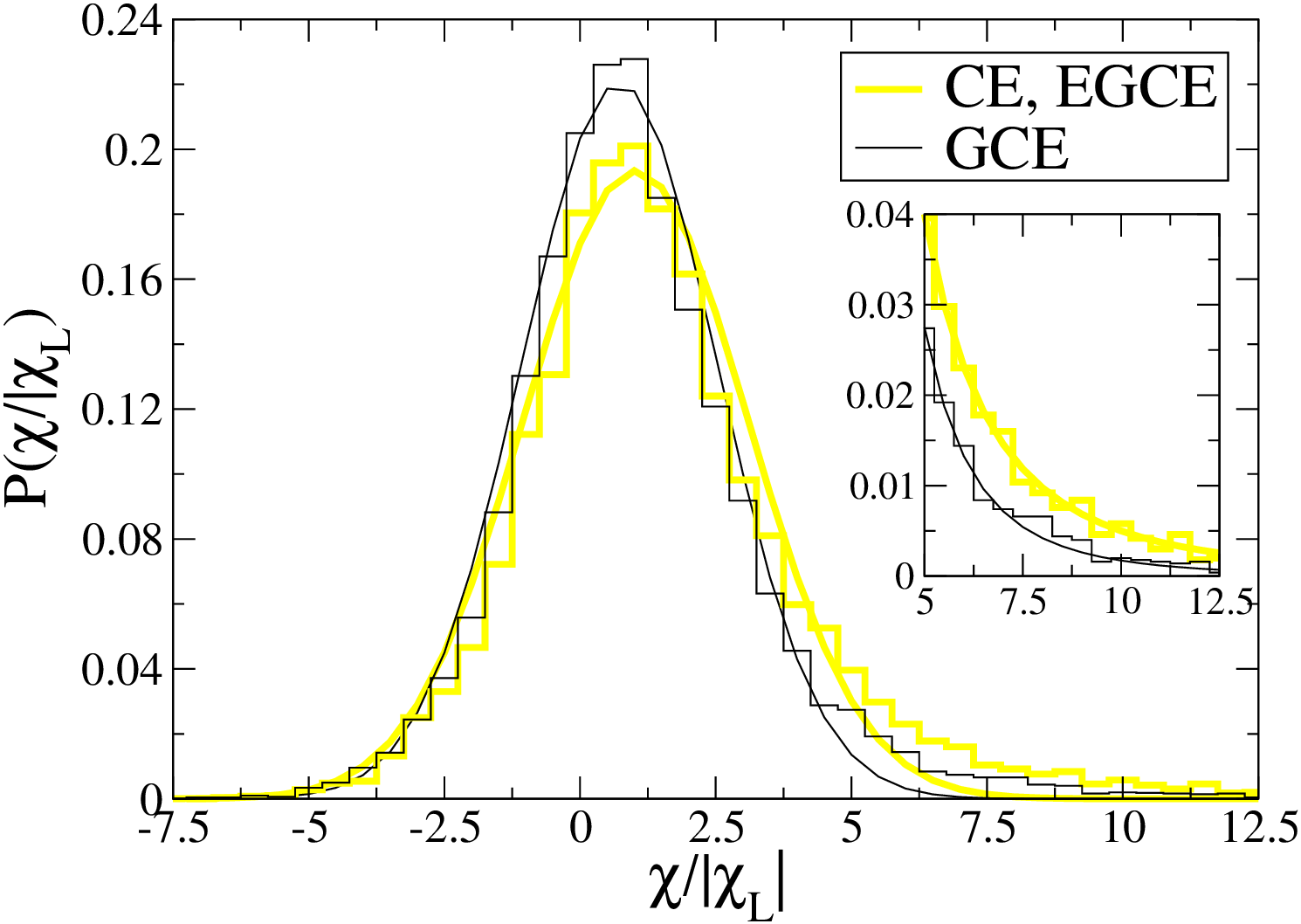}
\caption{\label{fig:distrib} The orbital susceptibility distribution of
a finite disordered lattice at zero temperature, for both the CE or
EGCE and the GCE for half-filling, with disorder values $W=2.0$ (upper
panel) and $W=4.0$ (lower panel). The tails of the distributions are
enlarged in the insets. Numerical results (histograms) are shown
together with fits (continuous curves). See the text for further
details.}
\end{figure}

The zero temperature susceptibility distribution in our systems is
shown in Fig.~\ref{fig:distrib}, for $W=2.0$ and $W=4.0$ and for both
the CE or EGCE and the GCE. In all the cases the results are for
half-filling. The numerically obtained distributions are shown
together with Gaussian fits. The tails of the distributions are
shown in the insets, together with fits to functional forms which we
derive below. We can see that all the distributions
are peaked at (or near) zero susceptibility, and are somewhat
asymmetric around their maximum. They are, however, fitted quite
well by a Gaussian near the peaks. On the other hand, The tails are
clearly non-Gaussian, decaying somewhat faster in the negative
direction and much more slowly in the positive direction. We note
that the susceptibility distributions for other fillings, where the
average susceptibility is different, are similar in form to
Fig.~\ref{fig:distrib}, but show the required tendency toward higher
or lower values.

We now compare the differences between the several cases shown.
We can see that the CE distribution is wider than the GCE distribution
for both disorder values, and has a stronger tendency for positive
values. In particular, the positive tail of the CE distribution is
larger than the corresponding tail in the GCE. When comparing the
two disorder values, note we have chosen a factor of 4 between the
scales of the two graphs. This factor is both as the ratio of the
widths of the Gaussian fits and a consequence of the expected
$W^{-2}$ dependence of the susceptibility moments. We can see that
after this rescaling the results for $W=2.0$ are similar to those
for $W=4.0$, although the distributions are somewhat wider in the
former case.

In two recent papers\cite{serota01,serota02} Serota has attempted
addressing the problem finding the zero temperature average
CE susceptibility and its fluctuations for diffusive systems. His
discussion was based on the expressions for the Larmor and van-Vleck
susceptibilities, Eq.~(\ref{eqn:larmor}) and Eq.~(\ref{eqn:vv}),
respectively. In order to take into account the above mentioned strong
cancellation between the Larmor and van-Vleck terms, he conjectured
that the only contribution surviving this cancellation is the largest
van-Vleck term composed of a matrix element and an energy difference
between the highest occupied level and the lowest unoccupied level,
which we will denote by $-$ and $+$ respectively. At zero temperature
one is thus left with the following expression for the zero
temperature susceptibility:
\begin{eqnarray} \label{eqn:serota}
\chi&=&\frac{1}{SH^2}{\left(\frac{e}{2mc}\right)}^2
\frac{ {\left|\langle +\left|\mathbf{A}\cdot\mathbf{{\hat p}}+
\mathbf{\hat p}\cdot\mathbf{A}\right|-\rangle\right|}^2 }
{\epsilon_{+}-\epsilon_{-}}.
\end{eqnarray}
Using an Gor'kov-Eliashberg\cite{gorkov65} like argument to
evaluate the squared matrix element semiclassically, and using the
Wigner distribution for the level spacing, Serota has given
expressions for the average and r.m.s. susceptibilities, which vary
as $|\chi_{{}_\mathrm{L}}|(k_F \ell)$ and $|\chi_{{}_\mathrm{L}}|
(k_F \ell) \left[\ln(E_C/\Delta)\right]^{1/2}$. These
expressions are similar to the corresponding expressions for
temperatures higher than the mean level spacing $\Delta$ which we
quoted in the introduction, if the temperature $T$ is replaced by
$\Delta$.

However, this approach is quite problematic in view of our previous
discussion of the importance of the gauge invariance of the
susceptibility. Eq.~(\ref{eqn:serota}) is clearly not gauge invariant,
since the Larmor--van-Vleck cancellation was taken into account a is too
crude fashion. As a result, Eq.~(\ref{eqn:serota}), can only give
positive susceptibility values, whereas our numerically computed
distribution, Fig.~\ref{fig:distrib} exhibits also negative values.

In spite of this problem, the approximation of Eq.~\ref{eqn:serota} is
still useful for treating
the tail of the susceptibility distribution. Our numerical simulations
indicate that anomalously large positive susceptibility values are
caused by small level separation between the highest occupied level
and the lowest unoccupied level. In those cases, the term in the
van-Vleck susceptibility involving these two levels is much larger than
all the other terms. Furthermore, in this case it is also approximately
gauge invariant, since a gauge transformation adds to the squared
matrix element in Eq.~(\ref{eqn:serota}) contributions smaller by a
factor of $(\epsilon_{+}-\epsilon_{-})/\Delta$ or less, which are thus
negligible for very close levels. Moreover, the average and r.m.s. of
the squared matrix element are seen in the numerics to be almost
independent of the energy separation for small separations, so that
the form of the tail of the susceptibility distribution is dominated
by the distribution of the energy denominator. This, in turn, can be
simply obtained from Wigner's distribution, according to which the
probability for small level separation varies linearly with the
separation. Thus, the susceptibility distribution varies as
$P(\chi) \mathrm{d}\chi \sim \chi^{-3} \mathrm{d}\chi$. This
dependence is fitted to the CE distributions in the insets of
Fig.~\ref{fig:distrib}, and agrees quite well with the numerical
results.

As previously noted by Serota, this distribution has a first moment but
not a second or higher moments, i.e., the r.m.s. susceptibility
diverges. Serota suggested regularizing this divergence by noting that
in any given experimental realization a finite magnetic field is used,
so that for very small energy separations the non-degenerate
perturbation theory used to obtain Eq.~(\ref{eqn:serota}) is
inapplicable. A degenerate perturbation theory must be used instead,
and this gives a finite magnetization.

All this is true, however, at zero temperature. At higher temperatures
Eq.~(\ref{eqn:serota}) should be multiplied by the factor
$(f(\epsilon_{-})-f(\epsilon_{+}))/2$, where $f(\epsilon)$ is the mean
level occupation. This factor tends to linearly zero for level
separation smaller than the temperature, and thus cuts off the
divergence caused by the denominator. For temperatures higher than
$\hbar \omega_c$ this factor, and not a finite magnetic field, is
dominant in keeping the susceptibility finite. In fact, using Serota's
estimate for the squared matrix element in Eq.~(\ref{eqn:serota}), the
r.m.s. susceptibility can be shown to vary as $|\chi_{{}_L}|(k_F \ell)
\left[\ln (\Delta/T)\right]^{1/2}$. This means that the logarithmic
temperature dependence of the r.m.s. susceptibility\cite{serota91}
does not saturate for temperatures smaller than the level spacing, but
persists to much lower temperatures, and, in the limit of zero
magnetic field, continues down to zero temperature, where the r.m.s.
susceptibility diverges.

This may serve to explain the strong temperature dependence of the
r.m.s. susceptibility seen in our numerical results. It also suggests
that the saturation of this effect at very low temperatures (less than
tenth of the mean level spacing) is only due to the finite ensemble
used, whereas in reality the r.m.s. susceptibility diverges as the
temperature becomes lower. To test this we compared the zero
temperature CE r.m.s. susceptibility (which is identical to the EGCE
susceptibility in this case) calculated numerically on ensembles of
different numbers of disorder realizations.
We observed, indeed, that the numerically calculated r.m.s.
susceptibility grows with the ensemble size, indicating it is really
divergent, and thus temperature dependent for all temperatures.

All the above refers to the CE or EGCE. Since the number of electrons
is fixed, the probability of having a small separation between the
lowest unoccupied level and the highest occupied level is given by the
Wigner distribution. In the GCE, however, there is a low probability
for the fixed chemical potential to fall between two close level,
and this probability is simply proportional to the separation itself
when the separation is small. The tail of the susceptibility
distribution thus varies as $\chi^{-4}$ instead of $\chi^{-3}$ in the
CE or EGCE. This dependence is fitted to the GCE distributions in the
insets of Fig.~\ref{fig:distrib}, and agrees quite well with the
numerical
results. Thus, the zero temperature r.m.s. susceptibility is expected
to be finite in the GCE (although higher moments are still divergent).
Numerical simulations have indeed shown that increasing the ensemble
size only helps smoothing the r.m.s. GCE susceptibility. This means
that the r.m.s. GCE susceptibility really saturates as a function of
temperature for low enough temperatures.

We thus arrive at a surprising conclusion: in contrast to what is
usually taken for granted, the r.m.s. susceptibility is different in
the CE (or EGCE) and in the GCE, diverging as a function of temperature
when approaching absolute zero in the former case, while tending to a
constant value in the former. Moreover, since the CE is qualitatively
similar to the EGCE at a lower temperature, the diverging temperature
dependence explains why we can see a noticeable difference between the
r.m.s. susceptibilities of the CE and the EGCE at very low temperatures.
(It is not very large, however, since the temperature dependence is
only logarithmic.)

To conclude this section we may remark that since the main difference
between the CE or EGCE and the GCE is connected with the events of
small level separation at the Fermi level, we can understand why the
average CE and EGCE susceptibilities show a temperature dependence
even for temperatures below the mean level spacing. However, the
average susceptibility is not divergent at zero temperature. This
means that we cannot refer only to the tails of the susceptibility
distribution, and the analysis becomes much more involved. In any
case, however, we
can conclude that the usual assumption of no temperature dependence
when $T<\Delta$ cannot be taken for granted even when treating the
average susceptibility.

\section{Interaction Effects} \label{sec:v}
It is generally accepted that interaction should have a strong effect
on the magnetic susceptibility. The interaction contribution to the
susceptibility is expected to be positive for a repulsive interaction
and negative for an attractive one
\cite{aslamazov75,altshuler83,serota91,jalabert99}. For diffusive
systems this effect is believed to be  insensitive to the ensemble
average chosen, and to have small fluctuations around its mean value.
In these systems, for weak interaction and temperatures higher than
the mean level spacing, this contribution varies as
$\lambda|\chi_{{}_\mathrm{L}}|(k_F \ell) \ln(T\tau/\hbar)$ where
$\tau$ is the elastic mean free time and $\lambda$ is the dimensionless
interaction constant (The logarithmic factor is correct in the very
low field regime $\Phi<\Phi_C=(T/E_C)^{1/2}\Phi_0$). For clean systems
this contribution varies as
$\lambda|\chi_{{}_\mathrm{L}}|(k_F L)^\alpha$ for some
geometry-dependent exponent $\alpha$ with a more complicated
temperature-dependent factor. In the following, however, we will
concentrate at zero temperature and repulsive interaction.

In order to avoid the heavy computation involved in exact
diagonalization or Hartree-Fock calculations, we will restrict
ourselves to weak short-range interaction, which can be treated
perturbatively to first order. We thus take as our model the usual
Hubbard on-site interaction (electronic spin is restored in this
section):
\begin{eqnarray} \label{eqn:hubbard}
U=U_H \sum_s {\hat n}_{s;\uparrow} {\hat n}_{s;\downarrow},
\end{eqnarray}
where $U_H$ specifies the interaction strength, and the arrow
subscripts denote spin direction. To first order in $U_H$ the change
in the system's energy at zero temperature is thus:
\begin{eqnarray}
\Delta E =
U_H \sum_{s} n_{\uparrow}(s) n_{\downarrow}(s),
\end{eqnarray}
where
\begin{eqnarray*}
n_{\sigma}(s) = \sum_{k=1}^{n_\sigma} |\psi_{k;\sigma} (s)|^2
\end{eqnarray*}
is the number of electrons with spin projection
$\sigma=\,\uparrow\,,\downarrow$ at lattice site $s$. In this later
expression $\psi_{k;\sigma}$ denotes the k-th single particle
eigenfunctions without interaction, with spin projection $\sigma$ (The
wave functions are, of course, spin independent; the spin index is used
only for clarity); and $n_\sigma$ is the total number of electrons
with spin projection $\sigma$ in the system (which is conserved by the
interaction). We will take these numbers as independent, as they may
be set different experimentally using an in-plane magnetic field. This
applies to the CE or EGCE; in the GCE the sum is over single-particle
levels with energies lower than the Fermi energy (which can again
depend on the spin direction).

The first-order interaction correction to the susceptibility is
$\Delta \chi=-S^{-1}\partial^2\Delta E/\partial H^2$. Since for
infinite clean systems $n_{\sigma}(s)$ is position independent both
with and without an applied magnetic, $\Delta E$ is field-independent
and $\Delta \chi$ vanishes. This is in agreement with previous
results \cite{aslamazov75}, according to which interaction corrections
to the susceptibility are of the third order in the coupling strength
($U_H$ in our case). On the other hand, for finite and/or disordered
systems, $n_{\sigma}(s)$ is non-uniform. Moreover, an applied
magnetic field breaks time-reversal symmetry and drives the electronic
density distribution to be more uniform. This makes $\Delta E$ smaller,
and thus $\Delta \chi$ is positive. This is again in accordance with
previous results \cite{aslamazov75,altshuler83,serota91,jalabert99}.

\begin{figure}
\includegraphics[width=8.5cm,height=!]{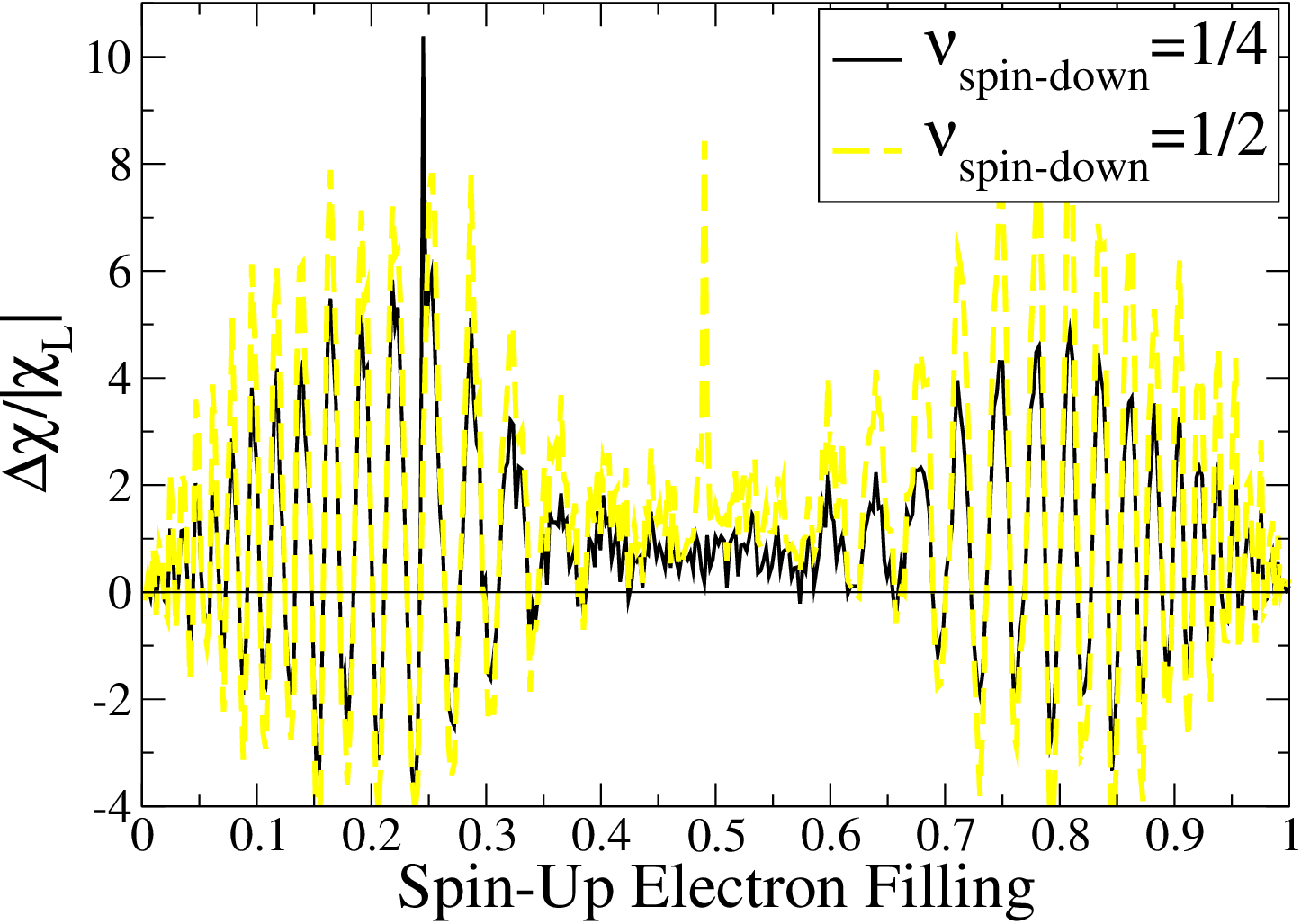}
\vskip 1.0cm
\includegraphics[width=8.5cm,height=!]{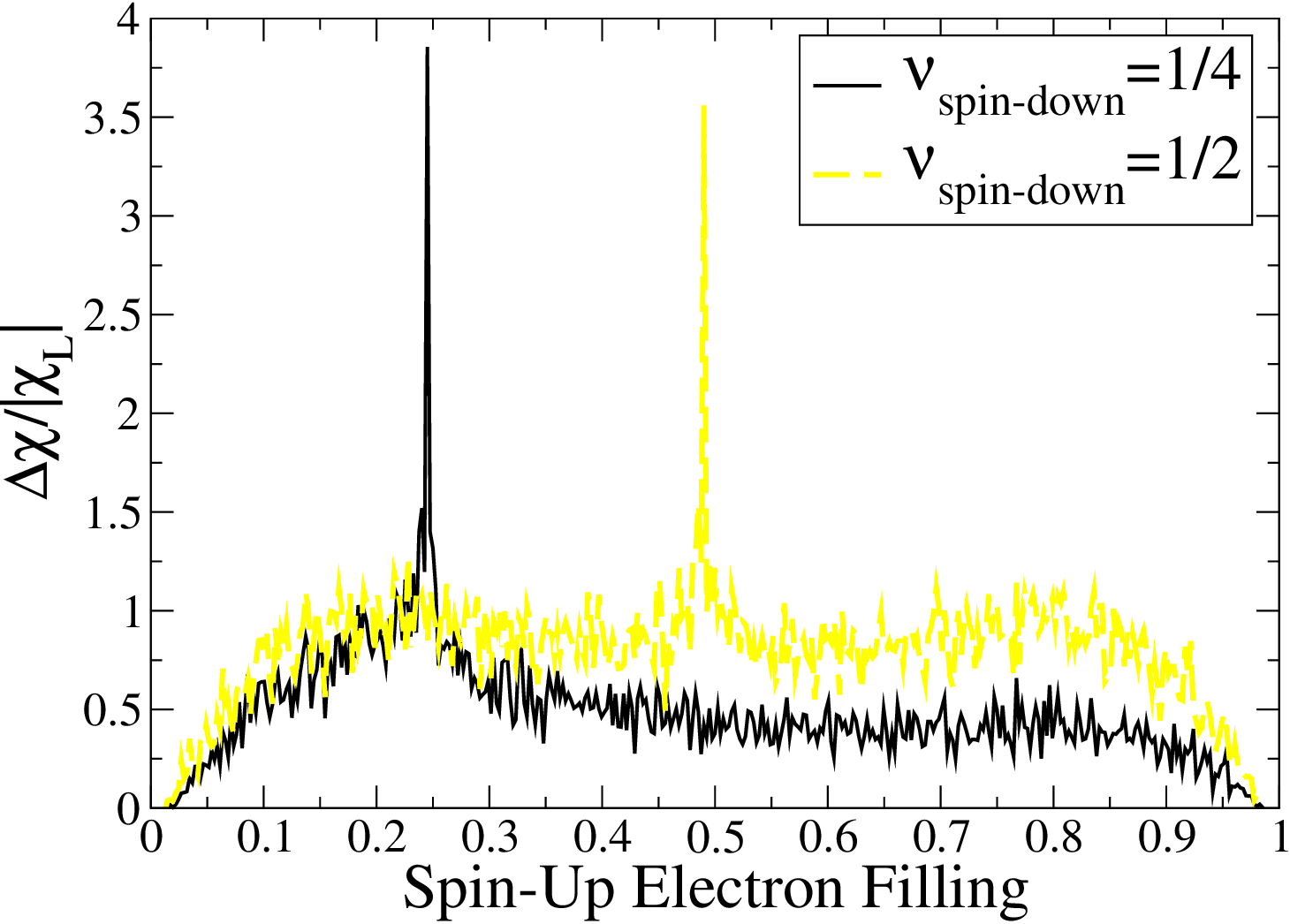}
\caption{\label{fig:inter} Corrections to the zero temperature
susceptibility of spin-up electrons due to their interaction with
spin-down electrons. The results are presented as a function of
filling of spin-up electrons for different fillings of spin-down
electrons in the CE (or EGCE), for both $W=2.0$ (upper panel) and
$W=4.0$ (lower panel).}
\end{figure}

We now turn to numerical results. Since one can show that the change
in the square of the absolute value of the wave-function amplitude
at a given point due to the applied magnetic field is quadratic in
the applied field, we have
\begin{eqnarray}
\Delta \chi =  -\frac{U_H}{S} \sum_s
\left[ \frac{\partial^2 n_{\uparrow}(s)}{\partial H^2}
n_{\downarrow}(s) + n_{\uparrow}(s)
\frac{\partial^2 n_{\downarrow}(s)}{\partial H^2} \right].
\end{eqnarray}
We may thus interpret the first term in the brackets as the correction
to the orbital susceptibility of the spin-up electrons due to their
interaction with spin down electrons, and similarly for the second
term. The first term was calculated numerically with our usual
parameters, and the results are shown in Fig.~\ref{fig:inter} for the
two values of disorder strength ($W=2.0$ and $W=4.0$). Since we are
using numerical differentiation, our results for $W=0$, which seemed
unreliable to us are not shown. For simplicity, We have taken the
interaction strength as our unit of energy, i.e., $U_H=t$.

As in the non-interacting case, for $W=2.0$ we can see ballistic
boundary-induced oscillations. The non-oscillating part of the
interaction correction shows the expected disorder dependence (varying
as $W^{-2}$ as $\ell$ should) and is size independent (This latter
result applies also to the oscillating part). An interesting feature
is the strong peak in $\Delta \chi$ when the filling of spin-up
electrons equals the filling of spin-down electrons. This can be
explained by treating the interaction in the Hartree-Fock manner, as
an effective single-particle potential. Obviously, the interaction
between two electrons in the same orbital single-particle level but
different spins is stronger than in the case where the two electrons
reside in two different orbital levels. Thus, the single-particle
energies of spin-up electrons up to the $n_\downarrow$-th level are
raised by a repulsive interaction more than the single particle
energies above the $n_\downarrow$-th level. Now, the expression for
the van-Vleck susceptibility, Eq.~\ref{eqn:vv}, contains at zero
temperature only energy
denominators between occupied and unoccupied levels, and is thus
dominated by the lowest unoccupied levels and the highest occupied
levels. When the filling of spin-up electrons differs enough from
the filling of spin-down electrons, the energies of these lowest
unoccupied levels and highest occupied levels are raised by
approximately the same amount by the interaction, so their differences
do not change very much and the corrections to the susceptibility are
small. However, when the filling of spin-up electrons is approximately
equal to the filling of spin-down electrons, the highest unoccupied
levels are raised by the interaction more than the lowest unoccupied
levels so that the energy differences become smaller and $\Delta \chi$
is enhanced, as we have seen in the numerical results.

To conclude this section we note that our numerical results indicate
that the r.m.s. value of $\Delta \chi$ is much larger than its average,
and that the average of $\Delta \chi$ in the GCE is significantly lower
than its CE (or EGCE) value, in contrast to what is usually expected.
It seems
that the distribution of $\Delta \chi$ at zero temperature shares with
the distribution of the non-interacting susceptibility the property of
having a long tail with diverging moments, causing a strong temperature
and statistical-ensemble dependence. This point, however, deserves
further investigation.

\section{Conclusions}
To conclude this paper, we summarize our main findings :

(i) For temperatures lower than the mean level spacing, we have shown
how the requirement of gauge-invariance causes a strong cancellation
between the Larmor and van-Vleck susceptibilities, and how it makes
using usual averaging procedures, like RMT, inapplicable for our
discussion. We have also seen that, contrary to what one usually
expects, the r.m.s. susceptibility depends strongly both on
temperature and statistical ensemble in this regime of low
temperatures. This was explained in term of the long tail of the zero
temperature susceptibility distribution, resulting in a diverging
r.m.s. zero temperature susceptibility in the CE or EGCE but not in
the GCE, and causing a logarithmic temperature dependence in the
former case for all temperatures down to zero. We believe this to be
responsible also to the strong temperature dependence of the average
CE and EGCE susceptibilities for this low temperature regime.

(ii) For temperatures higher than the mean level spacing our numerical
results generally agree with previous calculations. They, together
with analytical calculations show that the difference between the CE
and the EGCE, which was not considered so far in connection with
orbital susceptibility, is negligible for diffusive systems, and is
not the leading order contribution in the clean limit. However, the
meaning of the temperature where the EGCE susceptibility becomes equal
to the GCE value, and even more -- the temperature where the r.m.s.
susceptibility becomes equal to the average value, are not clear.
Especially in the latter case, identifying the temperature with the
Thouless energy is quite problematic.

(iii) Considering interaction between the electrons, we have given a
new interpretation of its first order contribution. We have
also demonstrated its dependence on in-plane magnetic field, and
explained this effect in terms of a single-particle picture.

\begin{acknowledgments}
We would like to thank B. Shapiro for useful discussions. Financial
support from the Israel Science Foundation is gratefully acknowledged.
\end{acknowledgments}

\appendix*
\section{Difference Between Ensembles in the Mesoscopic Regime}
In this appendix we apply the general formulation of Kamenev and Gefen
\cite{kamenev97} regarding the differences between ensembles to our
problem. They calculated $\delta F$, the disorder-averaged difference
between the canonical free energy $F$ and the expression
$\Omega+\mu N$, the Legendre transformation of the grand canonical
potential $\Omega$ calculated for system-independent chemical potential
$\mu$. The number of particles $N$ is thus sample-specific in the
grand-canonical case, and the chemical potential is set so that the
disorder and thermal averaged particle number will be equal to the
fixed particle number in the CE. Treating the regime of temperatures
higher than the mean level spacing
$\Delta$, they give the following expression, correct to the zeroth
order in the small parameter $T/\epsilon_F$:
\begin{eqnarray} \label{eqn:kamenev}
\delta F = \frac{\pi T}{2}
\sum_{\mathrm{w}=0}^{\infty} \frac{1}{(\mathrm{w}+1)!}
\left( \frac{\Delta}{T} \right)^\mathrm{w}
\int_{0}^{\infty} \mathrm{d}t \frac{t^{2\mathrm{w}}}{\sinh^2 \pi t}
{\widetilde K} \left( \frac{\Delta}{T} t \right), \nonumber \\
\end{eqnarray}
where
\begin{eqnarray*}
{\widetilde K}(t)=\frac{1}{\Delta}
\int_{-\infty}^{\infty} \mathrm{d}\omega
e^{-i\omega t/\Delta} K(\omega)
\end{eqnarray*}
is the Fourier transform of the two-level correlation function
$K(\omega)$. The $\mathrm{w}=0$ term is the difference between the
EGCE and the GCE, while the $\mathrm{w}=1$ is the leading order
difference between the CE and the EGCE. Only these terms will be
considered in what follows.

For temperatures lower than the Thouless energy $E_C$, the
magnetic-field dependent part of $K(\omega)$ can be approximated by
the zero-mode Cooperon contribution (since we are treating
temperatures higher than the mean level spacing, perturbative
analysis applies):
\begin{eqnarray} \label{eqn:K}
K(\omega)=\frac{\Delta^2}{2\pi^2}
\frac{\epsilon_H^2-\omega^2} { (\epsilon_H^2+\omega^2)^2 },
\end{eqnarray}
where $\epsilon_H\sim E_C (\Phi/\Phi_0)^2$ is the leading order shift
in the lowest Cooperon eigenmode due to the applied magnetic field
(valid for small magnetic fields, whose flux through the sample
$\Phi$ is smaller than the quantum flux $\Phi_0$).

Inserting this into Eq.~(\ref{eqn:kamenev}) we get for the difference
between the EGCE and the GCE the following expression, previously
derived by Oh \textit{et al.} \cite{serota91}, for the magnetization
per unit area:
\begin{eqnarray}
M_\mathrm{EGCE}-M_\mathrm{GCE} \sim
\frac{\Delta}{T} \frac{E_c \Phi}{\Phi_0^2}
\int_{-\infty}^{\infty} \mathrm{d}t
\frac{t^2}{\sinh^2(\pi t)} e^{-(\epsilon_H/T)t}. \nonumber \\
\end{eqnarray}
From this it follows that
\begin{eqnarray}
M_\mathrm{EGCE}-M_\mathrm{GCE} \sim \frac{\Delta}{\Phi_0}
\left\{ \begin{array}{cc}
\mbox{\large $ \frac{E_C}{T} \frac{\Phi}{\Phi_0} $},
& \Phi \ll \Phi_c; \\
\mbox{\large $\frac{\Phi_0}{\Phi} $},
& \Phi \gg \Phi_c; \\
\end{array} \right.
\end{eqnarray}
where $\Phi_c={(T / E_C)}^{1/2}\Phi_0$.
The difference between the CE and the EGCE is given by:
\begin{eqnarray}
M_\mathrm{CE}-M_\mathrm{EGCE} \sim
\left( \frac{\Delta}{T} \right)^2
\frac{E_c \Phi}{\Phi_0^2} \int_{-\infty}^{\infty} \mathrm{d}t
\frac{t^4}{\sinh^2(\pi t)} e^{-(\epsilon_H/T)t}. \nonumber \\
\end{eqnarray}
From this it follows that
\begin{eqnarray}
M_\mathrm{CE}-M_\mathrm{EGCE} \sim
\frac{\Delta^2}{T\Phi_0}
\left\{ \begin{array}{cc}
\mbox{\large $ \frac{E_C}{T} \frac{\Phi}{\Phi_0} $},
& \Phi \ll \Phi_c; \\
\mbox{\large
$ \left( \frac{T}{E_C} \right)^2
\left( \frac{\Phi_0}{\Phi} \right)^5 $},
& \Phi \gg \Phi_c. \\
\end{array} \right.
\end{eqnarray}
In our parameter regime we can see the the difference between the CE
and the EGCE is parametrically smaller than the
difference between the EGCE and the GCE. In fact, the former effect is
of the same order of magnitude as the contribution of higher order
perturbative corrections to $K(\omega)$ to the latter. This is in
accordance with our numerical results.

\end{document}